\documentstyle[twocolumn,aps,epsf,floats]{revtex}
\begin{document}

\twocolumn[\hsize\textwidth\columnwidth\hsize\csname %
@twocolumnfalse\endcsname


\title {Electronic structure of underdoped cuprates}
\author{Andrey V. Chubukov$^{1,2}$ and Dirk K. Morr$^{1}$}
\address{
$^{1}$Department of Physics, University of Wisconsin, Madison, WI 53706\\
$^{2}$P.L. Kapitza Institute for
Physical Problems, Moscow, Russia}
\date{\today}
\maketitle

\begin{abstract}
We consider a  two-dimensional
 Fermi liquid coupled to low-energy commensurate spin fluctuations.  
 At small coupling, the hole Fermi surface is large and 
centered around $Q =(\pi,\pi)$.
 We show that as the coupling increases, the shape of the quasiparticle 
Fermi surface and the spin-fermion vertex undergo a substantial evolution.
At strong couplings, $g \gg \omega_0$, where $\omega_0$ is the upper cutoff in
the spin susceptibility, the hole Fermi surface
consists of small pockets centered at $(\pm \pi/2, \pm \pi/2)$. Simultaneously,
the full
spin-fermion vertex is much smaller than the bare one, and scales nearly
linearly with $|q-Q|$, where $q$ is the momentum of the susceptibility.
At intermediate couplings, there exist both, a large hole Fermi surface
centered at $(\pi,\pi)$, and four hole pockets, but the
quasiparticle residue is small everywhere except for the pieces
of the pockets which face the origin of the Brillouin zone.  
The relevance of these results for recent
photoemission experiments in $YBCO$ and $Bi2212$ systems 
is discussed.
\end{abstract}
\pacs{PACS:  75.10Jm, 75.40Gb, 76.20+q} 
]

\section{introduction}

The physics of cuprate superconductors has been a very popular issue
for nearly a decade following the discovery of high temperature
superconductivity by Bednortz and M\"{u}ller in 1986~\cite{BM}.
 Over the past few years, 
it became increasingly clear to the ``high-$T_c$ community'' that the
mechanism of superconductivity is directly related to the unusual normal state
properties of the cuprates, particularly in the underdoped regime. 
The $^{63}$Cu spin-lattice relaxation rate and spin-echo decay
rate~\cite{Sli,BP}, uniform
susceptibility \cite{Johnson}, in-plane and c-axis resistivity
\cite{Ong,Iye},
all demonstrate a temperature and doping dependence which 
is different from the predictions of the conventional 
Landau Fermi liquid theory.
A behavior different from a conventional Fermi-liquid theory 
has also been observed in 
angle-resolved photoemission experiments
~\cite{Shen,LaRosa,Ding},
transport measurements~\cite{Hwang},
and optical experiments~\cite{Puchkov} on the underdoped cuprates.
Characterizing and explaining this behavior
is one of the major challenges presently facing the high-T$_c$
community.
A number of theoretical approaches to the cuprates have been proposed in recent years
\cite{And,ChaAnd,LeeWen,AIM,Varma,Pines}. One of the approaches to the cuprates, which we 
advocate in this paper,
is based on the assumption
that the physics of high-$T_c$ superconductors is governed by the close proximity
of superconducting and antiferromagnetic regions. Indeed, parent compounds
of cuprate superconductors ($La_2CuO_4$, $YBa_2Cu_3O_6$, $Sr_2CuO_2Cl_2$)
are antiferromagnetic insulators. Upon hole doping, the antiferromagnetic order
rapidly disappears and the system eventually becomes a superconductor. There
is plenty of evidence that medium-range magnetic correlations are still present
even at optimal doping (the one which yields the highest $T_c$). The most direct
evidence came from neutron measurements for optimally doped $La_{1-x}Sr_xCuO_4$ which
reported the observation of propagating 
spin-waves at scales larger than $J$~\cite{Hayden}.
Additional evidence for strong spin fluctuations even 
at optimal doping comes from
NMR and magnetic Raman experiments~\cite{Curro,Girsh}. 
                                      
The key element in the magnetic approach to cuprates is the assumption that
the dominant interaction between fermions is the exchange of spin
fluctuations~\cite{Pines}.
In this respect, the spin-fluctuation approach resembles a conventional BCS theory
with the only difference that the intermediate bosonic mode is a magnon
rather than a phonon. There is, however, one principal difference
between the magnetic and the phonon mechanisms - in the static limit, 
the exchange by spin fluctuations yields an effective interaction 
which is positive (i.e. repulsive) for all $q$:
\begin{equation}
\Gamma^{k+q,-k-q}_{k,-k} = g^2 \chi (q) \; ,
\label{inter}
\end{equation}
where $g$ is a coupling constant, and $\chi(q)>0$ is the static spin susceptibility.
In this situation,  the BCS-type
equation for the gap, 
\begin{equation}
\Delta_k = - g^2 \int \chi (k-k^{\prime}) \frac{\tanh
\frac{\epsilon_{k^{\prime}}}{2T}}{2\epsilon_{k^{\prime}}} \Delta(k^{\prime})\; d^2k
\label{bcs}
\end{equation}
cannot be satisfied by an $s-$wave type solution $\Delta_k = const$.
On the other hand, in the vicinity of an antiferromagnetic
region, the spin susceptibility is peaked at or near the
antiferromagnetic ordering momentum $Q=(\pi,\pi)$ such that Eq.(\ref{bcs}) 
relates $\Delta_k$ and $\Delta_{k+Q}$. One can then look for a
solution which satisfies $\Delta_k = - \Delta_{k+Q}$ in which case the overall
minus sign in Eq.(\ref{bcs}) disappears.
For the 2D square lattice and near half filling (where superconductivity has
been observed), this extra condition on the gap 
implies that it changes sign twice
as one goes along the Fermi surface, and vanishes along the Brillouin zone
diagonals. In the language of group theory, such a pairing state possesses
$d_{x^2 -y^2}$ symmetry ($\Delta_k \propto (\cos k_x - \cos k_y)$). At present,
there is a large body of evidence that this pairing state {\it is} the pairing
state in cuprates - the most direct evidence came from tunneling experiments on
$YBCO$~\cite{Kirtley}. Notice, however, that there still exist some experimental data which
at present cannot be explained within the $d-$wave scenario~\cite{Dynes}.
It has been argued in the literature that there is
an admixture of the $s-$wave component in the gap, which is very small for underdoped
materials, but increases as one moves into the overdoped
regime~\cite{Bob,Onellion}. This admixture
may be due to the fact that the materials contain
orthorhombic distortions which mix the $s$ and $d_{x^2 -y^2}$
representations~\cite{Annett}.

The effective interaction in Eq.(\ref{inter}) has been used to compute the
superconducting transition temperature in the Eliashberg formalism~\cite{MP,MS}.
 For $g \sim 0.6 eV$ inferred from resistivity measurements, and
$\chi (q)$ inferred from NMR data, these
calculations yield $T_c \sim 10^2K$ which is of the same order as the actual
value of $T_c$. The application of the same approach to underdoped cuprates,
however, yields a $T_c$ which steadily  increases as the system approaches
half-filling, simply because the spin susceptibility becomes 
more and more peaked at the antiferromagnetic momentum. 
Experimental data, however, show that
$T_c$ decreases when the system becomes more and more underdoped and eventually 
vanishes even before the system becomes magnetically ordered. 

\begin{figure} [t]
\begin{center}
\leavevmode
\epsffile{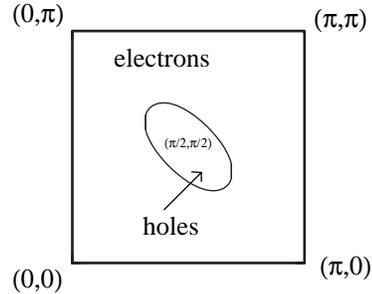}
\end{center}
\caption{A small Fermi surface consisting of a hole pocket centered 
around $(\pi/2,\pi/2)$.}
\label{pockets}
\end{figure}

Another piece of evidence concerning underdoped cuprates comes from recent
photoemission experiments which measure the electron spectral function 
$A(k,\omega) = (1/\pi) Im G(k, \omega)$ and
can therefore locate the Fermi surface in momentum space.
These experiments show that, at optimal doping, the hole 
Fermi surface is large, centered around $(\pi,\pi)$,
and encloses an area of about half the Brillouin zone,
in consistency with Luttinger's theorem~\cite{phopt}. For underdoped
cuprates, however, the data by the Stanford \cite{Shen,SS} 
and Wisconsin \cite{LaRosa} groups indicate that the
Fermi-surface crossing is present only for the momenta close to the zone
diagonal ($k_x \approx \pm k_y$).
No crossing has been observed near $(0,\pi)$
and symmetry-related points. Instead, the spectral function in these
regions of momentum space has a broad maximum at $100-200 meV$.
 This feature can be
interpreted as the quasiparticle peak in a strongly interacting
Fermi-liquid in which a substantial portion of the
 spectral weight is shifted
from the coherent peak into the incoherent continuum. Interpreted in this way,
the data indicate that the Fermi surface crossing 
in the underdoped cuprates
exists only for the directions close to the zone diagonal.
 This is consistent with the idea,
first put forward by Shraiman and Siggia~\cite{ShSi},  that the
hole Fermi surface in heavily underdoped cuprates 
consists of small hole pockets centered around $(\pi/2,\pi/2)$ and
symmetry-related points (see Fig.~\ref{pockets}).
Whether this idea is fully consistent with
the observations is still a subject of debate. If pockets exist, then one
should observe two Fermi surface crossings along the zone diagonal. 
Kendziora {\it et al.} argued that they did observe two crossings in their
measurements on $30K$ superconductor, and therefore  are able to
reconstruct the whole pocket~\cite{Onellion}.
 The data by Shen {\it et al.}~\cite{Sh} for the most heavily
underdoped superconductors also indicate
 that the maximum in the spectral function
first shifts to lower frequencies
 as one moves from $(0,0)$ to $(\pi/2,\pi/2)$, then
disappears before the $(\pi/2,\pi/2)$ point is reached, 
and then reappears and
moves towards higher frequencies as one moves from $(\pi/2,\pi/2)$ to
$(\pi,\pi)$. They, however, argued that it is difficult to determine
whether the reappearance of the 
peak between $(\pi/2,\pi/2)$ and $(\pi,\pi)$ is an observation of the second 
side of the pocket, or whether its reappearance possesses a structural 
origin~\cite{comm1}. 

Our interpretation of the photoemission data parallels the one made by
Kendziora {\it et al.} - we believe that there is substantial evidence that the
Fermi surface in the underdoped cuprates {\it does} consist of small hole
pockets. A natural question one may then 
ask is whether the formation of pockets is
related to the reduction of $T_c$. We will argue in the paper 
that this is, in fact, the case. Specifically, we will argue that in the doping
range where the Fermi surface evolves
 from a large one to a small one, 
vertex corrections substantially reduce the pairing interaction. 
How precisely this
evolution occurs is the subject of the present paper.
Notice that the evolution of the Fermi surface
cannot be continuous since the large and small
 Fermi surfaces are centered around different
points in momentum space. In fact, our results show that the Fermi surface
evolution should necessary include a
Lifshitz-type phase transition  
in which the topology of the Fermi surface changes. 

We conclude the introduction with the discussion of a microscopic model for the
cuprates. There are numerous reasons to believe that the low-energy 
properties of the cuprates are quantitatively captured by the effective 2D
one band Hubbard model \cite{SchrAnd,Bulut,Kampf}
\begin{equation}
{\cal H} = - \sum_{i,j} t_{i,j} c^{\dagger}_{i,\alpha} c_{j,\alpha} + U \sum_i
n_{i,\uparrow} n_{i,\downarrow}
\label{hub}
\end{equation}
Here $\alpha$ is a spin index, $n_{i,\sigma} = c^{\dagger}_{i,\sigma} c_{i,\sigma}$, 
and $t_{i,j}$ is a hopping
integral which we assume to act between nearest  and next-nearest neighbors
($t$ and $t^{\prime}$, respectively). By itself, this one-band model is a
simplification, since in ``first principles'' calculations one would start
with the three band model for the $CuO_2$ unit.  However, it is generally 
accepted
that at energies less than about $5 eV$ the hybridization between $Cu$ and $O$
orbitals is rather strong, and one can effectively describe the system by a
single degree of freedom per $CuO_2$ unit~\cite{oneband}.  
The on-site Coulomb repulsion $U$ which we use in Eq.(\ref{hub}) is close to
the actual charge-transfer gap between the unhybridized $Cu$ and $O$ bands which 
is about $2 eV$ .

Furthermore, the Hubbard model contains only electrons but no spins.
Spin fluctuations appear in this model as collective modes of fermions.
To obtain these modes, one has to sum up the 
RPA series in the particle-hole channel~\cite{Scal,CF,Sachdev}.
As a result of the summation, the product of the two fermionic Green's
functions is replaced by a spin susceptibility whose poles correspond to
spin fluctuation modes. The susceptibility thus obtained
takes the general form
\begin{equation}
\chi_{ii}({\bf q},\omega) = \frac{\tilde{\chi}_{ii}({\bf q},\omega)}{1-U
\tilde{\chi}_{ii}({\bf q},\omega)}
\label{chiq}
\end{equation}
in which $\tilde{\chi}_{ii}(\bf {q},\omega)$ is the irreducible particle-hole
susceptibility, and $i = x,y,$ or $z$.
Similarly, one can obtain the spin-fermion vertex (i.e.,
 the coupling between two fermions and one spin fluctuation),
by starting with the original
four-fermion Hubbard interaction term and 
combining one incoming and one outgoing
fermion into the RPA series.
One can then introduce an effective model in which 
fermions and spins are considered as independent degrees of freedom coupled by 
\begin{equation}
{\cal H}_{int} =
g \sum_{{\bf q,k},\alpha,\beta}~
c^{\dagger}_{{\bf k+ q}, \alpha}\,
{\vec \sigma}_{\alpha,\beta}\, c_{{\bf k},\beta} \cdot {\vec S}_{\bf -q}
\label{intham}
\end{equation}
where  $\sigma_i$ are the Pauli matrices, and the spin propagator 
(i.e. the Fourier
transform of the retarded spin-spin correlation function)  is the RPA
susceptibility $\chi
(q,\omega)$. The coupling constant $g$ is equal to $U$ in the RPA
approximation. In semi-phenomenological theories, this coupling, however, is
considered  as some effective input parameter \cite{MP}. The argument here is that
$g$ can be substantially renormalized from $U$ due to diagrams not included in
the RPA series.

The spin-fermion model is a convenient point of departure if one intends to study
the effects of spin fluctuations on the electronic spectrum. We point out, however, 
that this model neglects a direct fermion-fermion
interaction. We will see later in this paper that one actually needs this
interaction as a necessary term to restore the Ward identities.  

The spin-fermion model can be further simplified if one  
assumes some phenomenological form for the full spin susceptibility rather
than computing it in the RPA approximation.
This procedure can partly be justified. The point is
that only the imaginary part of the irreducible
particle-hole susceptibility comes from an integration near the Fermi surface
where the fermionic Green's function is known on general grounds. 
The computation
of the real part of $\chi (q,\omega)$, on the other hand,
involves an integration over 
regions far from the Fermi surface.  In these regions, the actual fermionic 
propagators can differ substantially from their values for the noninteracting 
fermions. This in turn implies that the RPA approximation may be 
insufficient, and a phenomenological form of $\chi$ with the parameters
deduced from the experimental data might be 
more appropriate. 

We now briefly discuss the phenomenological form of the susceptibility.
One can argue quite generally that in the vicinity of the 
antiferromagnetic phase, the full spin susceptibility behaves as~\cite{MMP}
\begin{equation}
\chi({\bf {q}}, \omega) =  \frac{\chi_{\bf Q}}{1 + ({\bf q} - {\bf Q})^2
\xi^2 - i \omega/\omega_{sf} - \omega^2\xi^2/c_{sw}^2 }
\label{chi}
\end{equation}
where $\chi_{\bf Q} = \alpha \xi^{2}$, $\xi$ is the correlation length
(measured in units of a lattice constant), $c_{sw}$
is the spin-wave velocity,
 and $\omega_{sf} = c^2_{sw}/2\xi^{2}\gamma$ where
$\gamma$ is a spin damping rate. The parameters in the phenomenological
susceptibility can be inferred from comparison to NMR and neutron scattering
data, as it was demonstrated by Millis, Monien and Pines~\cite{MMP}. The choice
of $\alpha$ is indeed arbitrary as it can be absorbed into the coupling
constant. Furthermore, as we said above, 
$\omega_{sf}$ can be computed in an arbitrary Fermi liquid provided that the
spin-fermion coupling is not too strong. 
In fact, a comparison of the 
calculated (within the above framework) and
measured $\omega_{sf}$ yields an information about the value of the
quasiparticle residue at the Fermi surface in optimally doped
cuprates~\cite{CM}. 

In the further discussion, we will consider both 
Hubbard and effective spin-fermion models. In the next section, we review
the large 
$U$ spin-density-wave approach for a magnetically ordered state at low doping, 
and show that
in the presence of long-range order, the full vertex, $g_{eff} (k)$, in fact
vanishes at the antiferromagnetic momentum $k=Q$. Simultaneously, the hole 
Fermi surface consists of four small pockets. We then argue that  
both of these features should survive when the system looses long-range
order, and disappear only at much larger doping concentrations when the system 
effectively looses its short-range magnetic order. 
In Sec~\ref{toy} we study the effective model
which reproduces the evolution from a large to a small Fermi surface as the
strength of the spin-fermion coupling increases. 
In Sec~\ref{vert} we compute the corrections  to the spin-fermion vertex and show
that these corrections substantially reduce the vertex for the range of
couplings when the Fermi surface has a small area. 
Finally, in Sec.~\ref{concl} we
state our conclusions and point to unresolved issues.

\section{Ordered state}
\label{sdw}
We begin our discussion with the magnetically-ordered state. Our emphasis 
here will be to obtain
 the form of the spin-fermion interaction vertex, and the shape
and the area of the quasiparticle Fermi surface. 

A straightforward way to study the ordered state is to apply a
 spin-density-wave (SDW) formalism~\cite{CF,SWZ}.
Suppose that the system possesses a commensurate antiferromagnetic order in
the ground state. Then 
the $z-$component of the spin-density operator, ${\vec S}(q) = (1/2) \sum_k 
c^{\dagger}_{k+q, \alpha}
{\vec \sigma}_{\alpha,\beta} c_{k,\beta}$, has a nonzero expectation value at
$q =Q = (\pi,\pi)$.
In the SDW approach, one uses the relation 
 $\langle \sum_k c^{\dagger}_{k+Q, 
\uparrow}c_{k,\uparrow} \rangle = 
-\langle \sum_k c^{\dagger}_{k+Q, \downarrow}c_{k,\downarrow}\rangle
 = \langle S_z \rangle$  to decouple the quartic term in Eq.(\ref{hub}).
 The truncated Hamiltonian is then 
diagonalized by a Bogoliubov transformation to new quasiparticle operators
\begin{eqnarray}
c_{k,\sigma} &=& u_k a_{k,\sigma} + v_k b_{k,\sigma}, \nonumber \\
c_{k +Q,\sigma} &=& sgn(\sigma) (u_k b_{k,\sigma} - v_k  a_{k,\sigma}),
\label{bogol}
\end{eqnarray}
where the fermionic Bogolyubov coefficients are 
  \begin{equation}
u_k =\sqrt{{1\over 2}\left(1+{\epsilon^{(-)}_k\over E_k}\right)},~~
v_k =\sqrt{{1\over 2}\left(1-{\epsilon^{(-)}_k\over E_k}\right)}.
\label{uv}
\end{equation}
Here $\epsilon^{(-)}_k = (\epsilon_k - \epsilon_{k+Q})/2$, $\epsilon_k = -2 t
(\cos k_x + \cos k_y) - 4 t^{\prime} \cos k_x \cos k_y$ is the single particle
dispersion, and $E_k = \sqrt{(\epsilon^{(-)}_k)^2 + \Delta^2}$,
 where $\Delta =
U \langle S_z \rangle$. In the cuprates,  
``first-principle'' calculations yield $t \sim 0.3eV$ and $t^{\prime} \sim
-0.06 eV$~\cite{Hyber}.
Note that $\epsilon^{(-)}_k$ only contains $t$, and, hence, the Bogoliubov 
coefficients do not depend on $t^{\prime}$.
The self-consistency condition for $\Delta$
has the simple form $1/U =  \sum_k 1/E^{-}_k$,
where the summation runs over all occupied states in the magnetic Brillouin zone.

The diagonalization using (\ref{bogol})
 yields two bands of electronic states (the conduction
and valence bands) with a gap $2\Delta \sim U$ in the strong coupling
limit ($U \gg t,t^{\prime}$):
  \begin{equation}
{\cal H}={\sum_{k\sigma}}' E^c_k a^\dagger_{k\sigma} a_{k\sigma} - E^v_k
b^\dagger_{k\sigma} b_{k\sigma},
  \label{add2}
  \end{equation}
where $E^{c,v}_k = E_k \pm \epsilon^{(+)}_k$, $\epsilon^{(+)}_k = 
(\epsilon_k + \epsilon_{k+Q})/2$, and
the prime restricts the summation to the magnetic Brillouin zone. 
We can further expand under the square root
and obtain $E^{c,d}_k = \Delta + J(\cos k_x + \cos k_y)^2 \mp 4 t^{\prime} \cos
k_x \cos k_y$, where $J = 4t^2/U$ is the magnetic exchange integral.

It is also instructive to present the expression for the fermionic Green's
function. It now has two poles at $\Omega = {\bar E}^{c,v}_k$  (${\bar E} =
E-\mu$, where $\mu$ is the chemical potential),
and vanishes at $\Omega = \epsilon_{k+Q}$. Namely, we have
\begin{equation}
G (k, \Omega) =  \frac{u^2_k}{\Omega - \bar{E}^c_k} + 
\frac{v^2_k}{\Omega - \bar{E}^v_k} ~=~
 \frac{\Omega - \bar{\epsilon}_{k+Q}}{(\Omega -\bar{E}^c_k) 
(\Omega -\bar{E}^v_k)} 
\label{sdw-like}
\end{equation}

\begin{figure} [t]
\begin{center}
\leavevmode
\epsffile{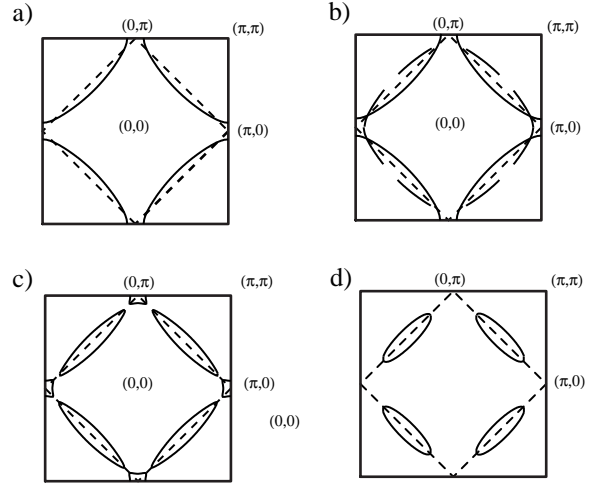}
\end{center}
\caption{Schematic
 evolution of the Fermi surface with the SDW gap $\Delta$ 
in the ordered phase of the $t-t^{\prime}$ Hubbard model. We set 
$t^{\prime}=-0.2t$. For $\Delta=0$ {\em (a)}
 the Fermi surface is large and centered around $(\pi,\pi)$.
 For $\Delta=0^+$ {\em (b)}
 additional 
pieces of the Fermi surface appear, which are the images of the original
 Fermi surface shifted by 
$(\pm \pi, \pm \pi)$. 
For intermediate $\Delta$ {\em (c)} one observes hole and electron 
pockets around 
$(\pi/2,\pi/2)$ and $(0,\pi)$, respectively.
 For large $\Delta$ 
{\em (d)} the Fermi surface just consists of 
hole pockets centered around $(\pi/2, \pi/2)$. The 
dashed line indicates the boundary of the magnetic Brillouin zone.} 
\label{fsord}
\end{figure}

\begin{figure} [t]
\begin{center}
\leavevmode
\epsffile{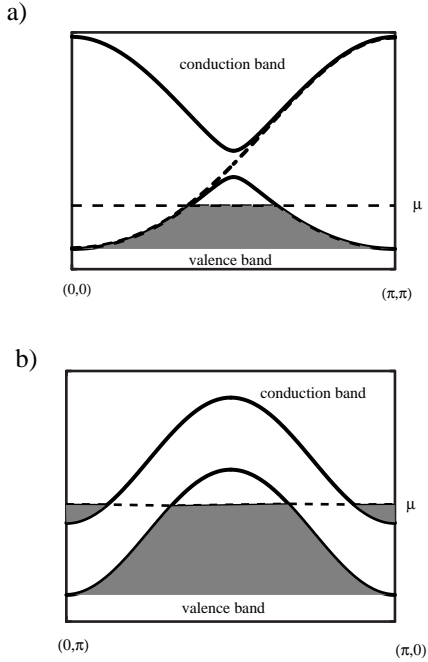}
\end{center}
\caption {The electronic spectrum along the $(0,0)-(\pi,\pi)$ 
direction {\em (a)} and the 
$(0,\pi)-(\pi,0)$ direction {\em (b)} for intermediate $\Delta$ 
(see Fig.\protect\ref{fsord}c). 
The shaded area corresponds to the filled 
electronic states. Note the presence of 
electron pockets around $(0,\pi)$ and $(\pi,0)$. The dashed line in 
{\em (a)} is the quasiparticle spectrum for $\Delta=0$.}   
\label{SDW_spectrum}
\end{figure}

Consider now how the Fermi surface evolves as $<S_z>$ and, hence, $\Delta$
increases. Suppose that we fix the doping concentration at 
some small but finite
level and increase $\Delta$ from zero to the strong coupling value 
$\Delta \sim U$. For $\Delta =0$, the Fermi surface for holes 
has the form shown in
Fig~\ref{fsord}a - it is centered at $(\pi,\pi)$ and 
encloses the area which is slightly larger than half of the Brillouin zone, in
accordance with the Luttinger theorem. 
Switching on  an infinitesimally small
 $\Delta$ doubles the unit cell in real space. As a result, the electronic
spectrum acquires  an extra branch which is just the image of the original
dispersion but  shifted by Q (see Fig.~\ref{SDW_spectrum}).
The same also happens with the Fermi surface - it  acquires extra
pieces because a Fermi surface crossing at $k$ necessarily implies one at $k+Q$
(Fig.~\ref{fsord}b). 
The subsequent evolution of the Fermi surface proceeds as  is
shown in Fig.~\ref{fsord}c,d. The details of this evolution are  model
dependent and therefore are not that relevant. An essential point is that,
at large $\Delta$, valence and conduction bands are well separated, and 
$u_k^2, ~v^2_k \approx 1/2$, i.e., $Z \approx 1/2$ for all momenta.
Since a summation of the electronic states 
with quasiparticle residue $Z=1/2$ over  
the full Brillouin zone  gives the same
result as a summation   of the states with  residue
$Z=1$ over  half of the Brillouin zone,  
nearly all states in the valence band are occupied, except for a fraction,
which is proportional to the doping concentration measured from half filling. 
This in turn means that the hole Fermi surface should be small and centered
around the minimum of $E^v_k$. For negative $t^{\prime}$ which we only consider
here, this minimum is located at $(\pi/2,\pi/2)$ \cite{CM1,Duf} (see Fig.~\ref{fsord}d). The
area of this small, pocket-like Fermi surface can be obtained either 
by modifying
Luttinger's arguments to account for
the existence of both, two poles and a vanishing numerator in
the Green's function, or by a direct computation of the particle density $N/V$.
Performing this computation, we obtain
\begin{equation}
x = S_{hole} - S_{elect}
\label{area}
\end{equation}
where $S_{hole}$ and $S_{electr}$ are the closed 
areas of hole pockets and doubly occupied electronic states (the latter appear
at intermediate stages of the Fermi-surface evolution, 
see Figs.~\ref{fsord}c and 
\ref{SDW_spectrum}b). At 
strong coupling
$(\Delta \sim U \gg t)$, the doubly occupied electron states disappear and we have
$S_{hole} =x$. 

We now proceed with the computations of the spin-fermion vertices. In the SDW
formalism, the low-energy spin fluctuations (magnons) are 
collective modes in the transverse spin channel. Consequently, they correspond
to the poles of the transverse spin susceptibility. The latter is given in the
SDW theory by the series of ladder diagrams. 
Consider first the strong-coupling limit at half-filling. Then
each ladder can only consist of conduction and valence fermions.   
Since the  unit cell is doubled due to the presence of the antiferromagnetic
long range order, we have two susceptibilities --- one with zero transferred
momentum, and one with the momentum transfer ${\bf Q}=(\pi,\pi)$.
The explicit forms of these susceptibilities are~\cite{CF}
\begin{eqnarray}
& &\chi^{\pm} (q,q,\omega) = - S~ 
\sqrt{1-\gamma_q\over {1+\gamma_q}}~\left[\frac{1}{\omega - \Omega_q +i\delta} 
- \frac{1}{\omega + \Omega_q -i\delta}\right],   \nonumber \\
& &\chi^{\pm} (q,q+Q,\omega) = -S~
~\left[\frac{1}{\omega - \Omega_q +i\delta} 
+ \frac{1}{\omega + \Omega_q -i\delta}\right]. 
\label{chi_sdw}
\end{eqnarray}
Here $\Omega_q = 4JS\sqrt{1-\gamma_q^2}$ is the magnon frequency.

Further, a sequence of bubble diagrams can be viewed as an effective
interaction between two fermions mediated by the exchange of a spin-wave.
The spin-wave propagators are
$i\langle Te_q(t) e^{\dagger}_q(0)\rangle_\omega =
(\Omega_q - \omega - i\delta)^{-1}$
and
$i\langle Te^{\dagger}_q(t) e_q(0)\rangle_\omega =
(\Omega_q + \omega - i\delta)^{-1}$,
where $e^{\dagger}_{q} (e_q)$ are the boson creation (annihilation)
operators, subindex $\omega$ implies Fourier transform, 
and the momentum $\bf q$ runs over the whole Brillouin zone.
A simple experimentation then shows that the form of the two susceptibilities
are reproduced if one chooses the following Hamiltonian for the interaction
between the original fermionic operators and the magnons~\cite{CF}: 
\begin{eqnarray}
H_{el-mag} &=& U~{\sum_{k}}^{\prime} {\sum_q} 
~\Big[\eta_q c^{\dagger}_{k+q,\alpha} c_{k,\beta} (e^{\dagger}_{-q} + e_q) \nonumber \\
&+&~{\bar \eta}_q c^{\dagger}_{k+q,\alpha} c_{k+Q,\beta} (e^{\dagger}_{-q} -
e_q)\Big]\delta_{\alpha,-\beta}.
\label{magn-ferm}
\end{eqnarray}
where 
\begin{equation}
\eta_q = \frac{1}{\sqrt{2}}\left(\frac{1-\gamma_q}{1+\gamma_q}\right)^{1/4},~~
{\overline\eta}_q = \frac{1}{\sqrt{2}}\left(\frac{1+\gamma_q}{1-\gamma_q}
\right)^{1/4}.
\end{equation}

Performing now the above Bogolyubov transformation, we obtain the Hamiltonian
for the interaction between the magnons and the conduction and valence
band fermions 
\begin{eqnarray}
H_{el-mag} &=&  
~{\sum_{k}}^{\prime} {\sum_q}~\Big[a^{\dagger}_{i\alpha k}a_{i\beta,k+q}e^{\dagger}_{q}
~\Phi_{aa} (k,q) \nonumber \\
&+& b^{\dagger}_{i\alpha k} b_{i\beta,k+q}e^{\dagger}_{q}
~\Phi_{bb} (k,q) + a^{\dagger}_{i\alpha k} b_{i\beta,k+q}e^{\dagger}_{q} ~\Phi_{ab} (k,q) 
\nonumber \\
&+& b^{\dagger}_{i\alpha k} a_{i\beta,k+q}e^{\dagger}_{q}
~\Phi_{ba} (k,q) +~{\rm H.c.}~\Big]~\delta_{\alpha, -\beta}.
\label{tranham}
\end{eqnarray}
The vertex functions are given by 
\begin{eqnarray}
\Phi_{aa,bb} (k,q) &=& U \Big[\pm \Big(
u_k u_{k+q} - v_k v_{k+q}\Big) 
\eta_q + \Big(u_k v_{k+q}  \nonumber \\
&-& v_k u_{k+q}\Big){\overline\eta}_q \Big], \nonumber \\
\Phi_{ab,ba} (k,q) &=& U~\Big[
\Big(u_k v_{k+q} + v_k u_{k+q}\Big) \eta_q \mp
\Big(u_k u_{k+q} \nonumber \\
&+& v_k v_{k+q}\Big) {\overline\eta}_q \Big]
\label{vertices1}
\end{eqnarray} 
In the strong coupling limit which we actually study now, they reduce to
\begin{eqnarray}
\Phi_{aa,bb} (k,q) &=& \left[\pm \Big(\epsilon^{(-)}_k+
\epsilon^{(-)}_{k+q}\Big)\eta_q + \Big(\epsilon^{(-)}_k-
\epsilon^{(-)}_{k+q}\Big){\overline\eta}_q
\right],
\nonumber \\
\Phi_{ab,ba} (k,q) &=& U~\left[
\eta_q \mp {\overline\eta}_q \right].
\label{vertices}
\end{eqnarray} 

We see that there are two types of vertices: $\Phi_{ab,ba}$ 
which describes the interaction
between conduction and valence fermions, and $\Phi_{aa,bb}$ which involves either
valence or conduction 
fermions. The first vertex is virtually not renormalized by 
Bogolyubov coefficients. However, at large $\Delta$, it involves high-energy
conduction fermions and therefore can be omitted.
The second vertex is the one relevant for the pairing mechanism since 
slightly away from half-filling, both valence fermions can 
simultaneously be on the Fermi surface. We see that this vertex
is substantially reduced at large coupling - it is of the order of the hopping
integral rather than of the order of $U$. Moreover, it is easy to verify
that $\Phi_{bb}$ in fact vanishes as the magnon momentum approaches $Q$.
For $q$ close to $Q$, we have, expanding Eq.(\ref{vertices}),
$\Phi_{bb} \propto t |q-Q|^{1/2}$. The vanishing of $\Phi_{bb}$ at $q=Q$ is
a consequence of the Adler principle:  
in the ordered state, magnons are
Goldstone bosons and their interaction with other degrees of freedom should vanish
at the ordering momentum to preserve the Goldstone mode to all orders in
perturbation theory.  Note in this regard 
that the 
vertex which includes conduction and valence fermions  should not be included 
into the corrections to the
spin propagator. Indeed, this vertex is already 
used in the RPA series which yields the spin susceptibility presented in Eq.(\ref{chi_sdw}),
and to include it into the corrections will just be double counting of the same
diagrams. From this perspective, the fact that $\Phi_{ab,ba}$ does {\it not} vanish for 
$q=Q$ is not a violation of the Adler principle.

We now construct the effective vertex for the spin-fermion model.
Since the factors $\eta_q$ and $\bar{\eta}_q$ are derived
from the spin susceptibility,  we have to absorb them into  $S_q$.
 Actually, only 
$\eta_q$ is relevant as it diverges at $q=Q$. 
Eliminating this factor in $\Phi_{bb}$, and substituting $U$ by the bare
coupling constant $g$, 
we find that the effective spin-fermion interaction behaves as 
\begin{equation}
g_{eff} = g~\left(u_k u_{k+q} - v_k v_{k+q}\right)  \propto t |q-Q|.
\label{ward}
\end{equation}
This result was first  obtained by Schrieffer \cite{Schr}.
We see that as long as one can keep the conduction fermions far away from the Fermi
surface (which is the case at strong couplings),
the effective pairing interaction between fermions is substantially
reduced by a vertex renormalization \cite{comm5}.
Consider now the full static interaction between fermions
mediated by spin fluctuations. It  is given by $\Gamma (q) =
 (g_{eff} (q))^2~\chi(q)$. Since $\chi(q) \propto (q-Q)^{-2}$ and
$g_{eff} (q) \propto |q-Q|$, $\Gamma (q)$ tends to 
a {\it positive} constant as $q$ approaches $Q$.
On the other hand, however,  we would need a substantial
enhancement of the interaction near $Q$ for the appearance of $d-$wave 
superconductivity. We see, therefore, that the presence of a long-range order
actually excludes the occurrence of $d-$wave superconductivity at large $U$.  
In the next section we will study in detail what happens when the sublattice 
magnetization 
decreases and the mean-field gap, $\Delta=U<S_z>$, becomes smaller than the 
fermionic bandwidth despite a large $U$.
At this point, however,
we merely emphasize the correlation between the relevance of vertex
corrections and the shape of the Fermi surface, namely, that
when  the Fermi surface consists
of small hole pockets, vertex corrections are relevant and the full vertex is
much smaller than the bare one.

\subsection{Fluctuation effects}
\label{fleff}

Our next goal is to study how fluctuations modify this simple mean-field
picture. We will show that 
the mean-field  form of the quasiparticle dispersion 
remains  qualitatively (and even partly quantitatively)
correct even at strong couplings 
though there is a substantial shift of the spectral weight from the
coherent part of the spectral function
 to the incoherent part which stretches upto frequencies of a few $t$.
 At the same time,
the relation $\Delta = U<S_z>$ does not survive beyond mean-field theory, and 
the strong-coupling, SDW-like form of the quasiparticle dispersion 
with a gap between two bands holds even when
the system looses long-range antiferromagnetic order. More specifically, we will
show that there exist two different self-energy corrections. 
One self-energy correction is strong but nonsingular.
It gives rise to a shift of the spectral weight towards the incoherent part of the spectral 
function, but
preserves coherent excitations on the scale of $J$. The second correction 
renormalizes the gap but does not
change the form of the quasiparticle Green's function. 
This correction  is however singular for vanishing $<S_z>$ and  breaks 
the relation between $\Delta$ and $<S_z>$ such that the gap survives and remains 
of $O(U)$ when $<S_z>$ vanishes.  
To perform a perturbative expansion around the mean-field SDW state, we need an
expansion parameter. Obviously, an expansion in $U$ will not work as we assume
that $U \gg t$. There exists, however, a formal way to make the mean-field theory
exact - one should extend the original Hubbard model to a large number of
orbitals at a given site, $n_c =2S$, and perform an expansion in
$1/S$~\cite{CM1,AH}. The
$S=\infty$ limit corresponds to the 
mean-field solution, and all corrections are in powers of $1/S$. The
mean-field result for the spectrum is the same as in Eq.(\ref{add2})
 apart from the
extra factor of $2S$ (we also have to redefine ${\bar t} = t/2S$ as $t$ now
scales with $S$).

\begin{figure} [t]
\begin{center}
\leavevmode
\epsffile{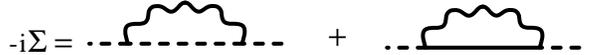}
\end{center}
\caption {The lowest order self-energy corrections for the valence 
fermions. The solid and dashed lines are
 the bare propagators for the conduction and 
valence fermions, respectively. The wavy line describes transverse spin 
fluctuations.}
\label{se}
\end{figure}

Consider now the lowest-order self-energy correction to the propagator of
the valence fermions. To first order in $1/S$, there are two self-energy diagrams,
one  involves only 
valence fermions, and the other
involves both, valence and conduction fermions (Fig.~\ref{se}).
In the first diagram,  the vertex is reduced
from $U$ to $O(t)$ and vanishes at $q=Q$. Numerically, however, this diagram yields large 
corrections since the large term, $\Delta \sim US$, in the energy denominator
is cancelled out because both fermions are
in the valence band. The denominator is then of the order of $O(JS)$.
As a result, the self-energy correction from this diagram scales as
$O({\bar t}^2/JS) = O(U)$ and obviously contains some dispersion at this 
scale~\cite{CM1}. At the same time, the bare dispersion is at the scale of
$JS$. We see that one    
clearly needs more than just smallness of $1/S$ to make perturbation 
theory work -
 the ratio $U/JS$ should be small too. This last assumption is  not
justified for high-$T_c$ materials where  $(U/JS)$ is
typically of the order of $\sim 20-30$. 

We recently considered this diagram in detail and 
performed  self-consistent calculations of the electronic
dispersion assuming that
$1/S \ll 1$ but also $U/JS \gg 1$~\cite{Dirk1}. One can show that in this limit
one can neglect vertex corrections but has to include the whole series of
rainbow diagrams for self-energy corrections~\cite{KaneLeeRead}. This corresponds 
to the insertion of the full Green's function into the internal fermionic line 
in Fig.\ref{se}. 
The self-consistency equation for $G(k,\omega)$ then takes the following form:
\begin{eqnarray}
G^{-1}(k,\omega)&=& \omega+\epsilon_k^{MF} \nonumber \\
& & \; -\int{ d^2q \Psi(k,q) G(k+q,\omega+\Omega_q) }
\label{sc}
\end{eqnarray}
where 
$$\epsilon_k^{MF}=\Delta+2JS (\cos k_x + \cos k_y)^2 - 4 t^{\prime} 
\cos k_x \cos k_y $$
is the mean-field dispersion and 
\begin{eqnarray*}
\Psi(k,q) &=& 32\bar{t}^2S \Big[\nu_k^2+\nu_{k+q}^2-2\nu_k \nu_q \nu_{k+q} \\
&+& \sqrt{1-\nu_q^2}(\nu_{k+q}-\nu_k)\Big]/\sqrt{1-\nu_q^2}
\end{eqnarray*}
Note that Eq.(\ref{sc})
is an integral equation in both momentum and frequency space.

The form of the quasiparticle dispersion in the ordered phase is interesting 
in its own 
because of recent photoemission data on $Sr_2CuO_2Cl_2$~\cite{Wells,Onel2}.
 These data show that (i) the valence band 
dispersion is isotropic near the  top of the 
band which is at $(\pi/2,\pi/2)$, and
relatively well described by the
mean-field formula with $t^{\prime} = -0.5J$, but (ii) coherent excitations
exist only near $(\pi/2,\pi/2)$. As soon as one 
deviates from this point, the width of the quasiparticle peak rapidly
increases, and at the scale of $2J$, the excitations become mostly incoherent.
Monte-Carlo and finite cluster 
simulations in the insulating phase have similarly 
demonstrated that  there exist coherent excitations at scales smaller than
$2J$, but also incoherent excitations which stretch upto a
few $t$~\cite{Dag,Hanke}.

\begin{figure} [t]
\begin{center}
\leavevmode
\epsffile{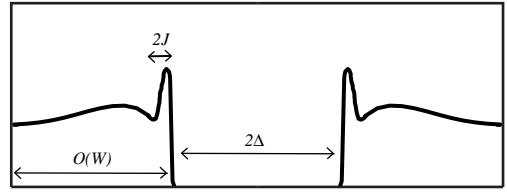}
\end{center}
\caption {Schematic form of the total DOS resulting from
 the self-consistent solution of Eq.\protect\ref{sc}. $W$ is the 
 bandwidth for free fermions.}
\label{dos}
\end{figure}

Our analytical results are consistent with the experimental data and with the
numerical
simulations. We first discuss the critical value of $\omega$ for which one 
first observes a finite imaginary part of $G(k,\omega)$, or equivalently, 
the onset of a finite density of states (DOS)~\cite{comm2}. On a mean-field
level, this happens on the scale of $\omega + \Delta = O(JS)$.
The analysis of Eq.(\ref{sc}) shows, however, that the onset of a DOS actually 
occurs at the much larger scale of $\omega + \Delta = O({\bar t} \sqrt{S})$. 
At these energy scales, the frequency shift due to  
$\Omega_q$ on the r.h.s.~of Eq.(\ref{sc}) can be neglected, and we just have to
solve an integral equation in momentum space. Doing this by standard means, we
obtain that a nonzero DOS appears at 
$$ \omega_{cr} + \Delta = \pm 4.2 \bar{t}\sqrt{2S}. $$
In the vicinity of $\omega_{cr}$, the DOS behaves as $N(\omega) \sim
\sqrt{|\omega-\omega_{cr}| }$. We see, therefore, 
that the incoherent part of the Green's
function stretches upto scales comparable to the original fermionic bandwidth.\\
Another important issue is the form of the quasiparticle Green's function
at $|\omega_{cr} - \omega| \leq JS$,  where one can show that 
the contribution
from the magnon dispersion in fact cannot be neglected. In addition, photoemission 
experiments have shown strong evidence for coherent fermionic excitations in this region.
We therefore assumed
the following trial form of the fermionic  Green's function
in the vicinity of $\omega_{cr}$
\begin{equation}
G(k,\omega) = {Z_k \over \omega - \omega_{cr} +A(k)-i \gamma (\omega - 
\omega_{cr})^2 }
\label{ans}
\end{equation}
Performing self-consistent computations in this region, we found~\cite{Dirk1}
 that the quasiparticle residue is indeed small, $Z \sim O(J\sqrt{S}/t) \ll 1$.
Analogous result has been obtained by Kane, Lee and Read~\cite{KaneLeeRead}. 
At the same
time, $A(k)$ is zero right at 
$(\pi/2,\pi/2)$ and scales quadratically with the deviation 
from this point. The scale for the coherent dispersion is $JS$ - the
same as for the mean-field solution. Even more, we recovered the
isotropic dispersion near $(\pi/2,\pi/2)$, namely  
 $A(k)=2JS \tilde{k}^2, \quad {\tilde{k}}=(\pi/2,\pi/2)-{k}$, 
and obtained the same bandwidth $2J$ as in the experimental data.
For the damping rate we found $\gamma \sim 1/JS$. 
It follows from our results that there exists just one typical energy 
scale $JS$ which separates
coherent excitations at smaller energies and incoherent excitations  at
higher energies (see Fig.\ref{dos}).
It is essential that this holds in the strong-coupling limit $t \gg J
\sqrt{S}$ when the quasiparticle residue $Z\ll 1$, and the mean-field dispersion
is completely overshadowed by self-energy corrections. In this respect, the
existence of coherent excitations at the scales of $JS$ turns out to be an
intrinsic property of the Hubbard model and not the result of a mean-field
approximation. Moreover, since the vertex does not depend on $t^{\prime}$, the
strong-coupling results are independent of the $t^{\prime}/J$ ratio, contrary
to the mean-field dispersion~\cite{comm3}.  Notice also that  
the dominant  corrections due to valence-valence vertex 
come from short-range spin fluctuations. The contribution
from long-range fluctuations with $q \approx Q$ is reduced
due to a vanishing vertex at $q=Q$. In other words, the valence-valence diagrams
are certainly non-singular. Not surprisingly then, these corrections can
reduce the coherent part of the quasiparticle dispersion but cannot 
qualitatively change the mean-field description. For the sake of simplicity,
we will just neglect these corrections for the rest of the paper.

\begin{figure} [t]
\begin{center}
\leavevmode
\epsffile{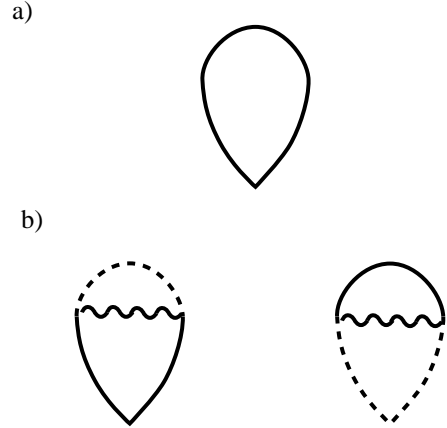}
\end{center}
\caption {The loop diagrams for $<S_z>$. 
The diagram in 
{\em (a)} represents $<S_z>/S$ on the mean-field level while those in
{\em (b)} correspond to the lowest 
order corrections to $<S_z>/S$ due to the exchange of transverse fluctuations.}
\label{Sz}
\end{figure}

We will now discuss the self-energy correction which involves both conduction and
valence fermions. Here the vertex is of the order of
$U$, but the energy denominator  
scales as $US$. As a result, one obtains a
momentum-independent correction to the gap, and
also (doing expansion in $J/U$ and in $\omega/U$, where $\omega$ is the 
external frequency) a correction to the dispersion on the scale of
$O(J)$. The latter is nothing but the conventional 
$1/S$ correction to the mean-field dispersion which is
obviously small at large $S$. In other words, the inclusion of the 
self-energy diagram with valence and conduction fermions leaves the
structure of the quasiparticle Green's function unchanged, but renormalizes the 
gap.  It is essential, however, that the gap renormalization can be made
singular when $<S_z>$ vanishes. The easiest way to see this
is to consider what happens at infinitesimally small temperatures in $2D$.
The Mermin-Wagner theorem tells us that immediately as the temperature becomes 
finite,
the long-range order should disappear. In perturbation theory, the onset of
this effect is the appearance of  logarithmically divergent correction which
scales as $T \log L$ where $L$ is the system size. We evaluated the self-energy
correction at $T=0^+$ and obtained
\begin{equation}
\Sigma = - U~ {1 \over N} \sum_q { 1 - \sqrt{1 - \nu^2_q} \over  \sqrt{1-
\nu^2_q}}~(1 + 2 n^b_q)
\label{corrsz1}
\end{equation}
where $n^b_q = (exp (\Omega_q/T) -1)^{-1}$ is the Bose distribution function. 
At any finite $T$, $\Sigma$ is logarithmically divergent:
$\Sigma \propto T \int d^2 q /q^2 \sim T \log L$. If this were the only correction,
the mean-field description would be completely destroyed. 
However, there exist another logarithmically divergent diagram which cancels
the contribution from $\Sigma$. The point is that the mean-field expression 
for the gap $\Delta_{MF}
= U <S_z>$ contains the {\it exact} sublattice magnetization.  On the
mean-field level, one has (at large $U$), $<S_z> \approx S$. 
When $T$ becomes nonzero, $<S_z>$ also acquires a
logarithmically divergent correction which involves the valence-conduction 
vertex (see Fig.~\ref{Sz}b).
In analytical form, we have
\begin{equation}
\delta (<S_z>) = - {1 \over N} \sum_q {1 - \sqrt{1 - \nu^2_q} \over \sqrt{1-
\nu^2_q}}~(1 + 2 n^b_q)
\label{corrsz}
\end{equation}
Combining these two terms, we find that the logarithmically divergent
corrections to  $\Delta$ 
 are cancelled out. As a result, $\Delta$ remains finite
even when the long-range order disappears. 
Moreover, a comparison of Eq.(\ref{corrsz1}) and Eq.(\ref{corrsz})
shows that the gap remains {\it exactly} equal to $US$ in the large $U$ limit. 

\begin{figure} [t]
\begin{center}
\leavevmode
\epsffile{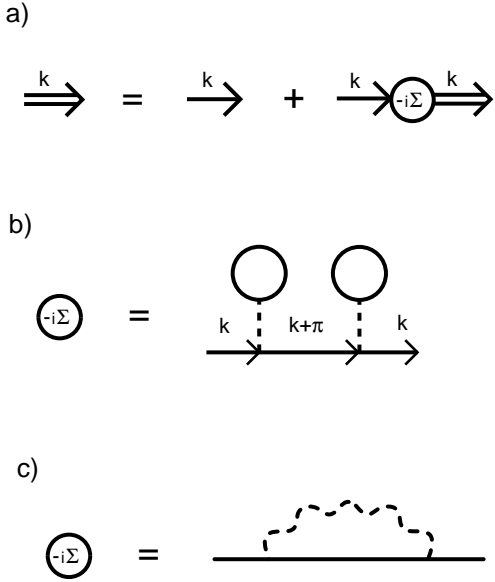}
\end{center}
\caption {{\em (a)} The Gorkov equation for the normal Green's function
for the SDW state. 
 The single and double solid lines 
correspond to the bare and full fermionic propagators. 
{\em (b)} The self-energy diagram which includes two
anomalous loops
The straight dashed line represents the 
interaction term $U$. {\em (c)} Formal representation of the same diagram
as in {\em (b)} with the exchange of a $\delta-$functional longitudinal spin 
fluctuations.}
\label{Gorkeq}
\end{figure}

By itself, this result is not surprising since at large $U$, the valence and 
conduction bands
are nothing but the lower and upper Hubbard bands, respectively. 
In the near atomic limit,
the gap between the two bands should be equal to $U$ (for $S=1/2$) 
upto corrections of $O((t/U)^2)$
independent of whether or not the system is ordered. We extended our 
calculations
to analyze the form of the full quasiparticle 
Green's function and found that not only $\Delta$ but also
$G_{full}(k ,\omega)$ is free from logarithmical singularities and
therefore does  not undergo sharp changes when one looses long-range order due to singular
thermal fluctuations.  
In other words, we found that the SDW structure of the electronic
states with two coherent bands  
separated by a large gap $\approx U$  survives
when the system looses long-range order. We remind that the key features of 
the SDW state at large $U$
 are the small Fermi surface
and a strong reduction of the pairing vertex. We see that both features may exist
even without a long-range order. The interesting issue 
then arises how the system evolves with doping
in the paramagnetic state 
and how it eventually restores a large Fermi
surface and a momentum independent vertex function.
Notice that without strong magnetic fluctuations, the two bands which emerge from 
the Hubbard levels are likely to be mostly incoherent, and there 
also appears, at finite $t$, a coherent band at about $U/2$. This was found in 
the infinite-$D$ studies of the Hubbard-model \cite{Roz}. 
Before we proceed to the discussion of the Fermi surface evolution, we consider
another, complimentary approach to the ordered
phase, introduced by Kampf and Schrieffer \cite{KS}. 
They demonstrated that one can 
obtain a mean-field SDW solution in the spin-fermion model 
without introducing a condensate, but by assuming that the longitudinal spin
susceptibility has a $\delta-$functional peak at zero 
frequency and momentum transfer
$Q$: $\chi_l (q,\omega) = (1/4)~\delta (\omega) \delta (q-Q)$.
To recover the mean-field SDW result within the  
spin-fermion model one has to compute 
the lowest-order self-energy diagram which includes {\it only} 
the exchange of  
longitudinal fluctuations. This diagram is shown in Fig.\ref{Gorkeq}c.
We then obtain
\begin{equation}
\Sigma (k, \omega) = \frac{g^2}{4}~\frac{1}{\omega - \bar{\epsilon}_{k+Q}}
\end{equation}
where, as before, $\bar{\epsilon}_k = \epsilon_k - \mu.$
Substituting this diagram into the Dyson equation, one immediately
recovers the mean-field SDW result for the spectrum with $g$ being the 
equivalent of $U$. 

\begin{figure} [t]
\begin{center}
\leavevmode
\epsffile{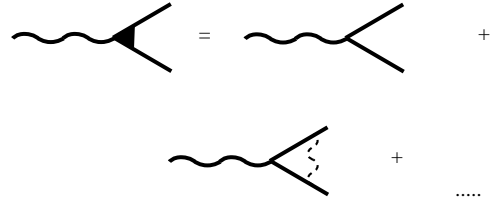}
\end{center}
\caption {The diagrammatic representation for the
full vertex (filled triangle)
 between transverse spin fluctuations and fermions in the ordered state.
The solid and dashed wavy lines correspond to the 
transverse and longitudinal spin 
fluctuations, respectively. The solid straight lines 
are the bare fermion Green's functions. Note that the momenta of the 
incoming and outgoing fermions differ by $Q=(\pi,\pi)$.}
\label{figsovert}
\end{figure}

Furthermore, one can also reproduce the correct form of the vertex between 
fermions and transverse spin fluctuations. The bare vertex is equal to $g$. 
To restore the correct mean-field form of the full vertex, 
one has to include the 
second-order vertex renormalization due to longitudinal  
spin fluctuations.
The second order vertex correction is given by the diagram in 
Fig~\ref{figsovert}.
The evaluation of this diagram is straightforward, and we obtain
\begin{equation}
\delta g = - \frac{g^3}{4}~\frac{1}{(\omega - {\bar \epsilon}_{k+Q}) (\omega+\Omega - 
{\bar \epsilon}_{k+Q+q})}      
\end{equation}
where $q$ is the magnon momentum, and $\omega$ and $\Omega$ are the external fermionic and 
magnon frequencies, respectively.
Combining this result with $g$, we find the total vertex in the form
\begin{equation}
g_{tot} = g~ \frac{(\omega - {\bar \epsilon}_{k+Q}) (\omega + \Omega - 
{\bar \epsilon}_{k+q+Q}) - (g/2)^2}{ (\omega - {\bar \epsilon}_k) (\omega + \Omega - 
{\bar \epsilon}_{k+q+Q})}      
\label{fullG}
\end{equation}
For the vertex which involves only valence fermions, we have 
at the mass surface, $\omega = -E^v_k,~\omega + \Omega = - E^v_{k+q}$. Substituting
this into Eq.(\ref{fullG}) and doing elementary manipulations, we obtain 
the same expression as in Eq.(\ref{ward}). This result has also been 
obtained by Schrieffer \cite{Schr}.

One may wonder what happens with the higher-order self-energy 
and vertex corrections.
In fact, they simply do not exist. The reason is that the use of the
$\delta-$functional form of the longitudinal susceptibility is just 
a way to reexpress the mean-field decoupling without formally introducing the
condensate. Specifically, one can rewrite the Gorkov equations for the SDW state
by introducing the condensate, but  
without formally introducing the anomalous Green's function, as is shown in
Fig.~\ref{Gorkeq}.
 The self-energy diagram obtained in this way is shown in
Fig.~\ref{Gorkeq}b. We can now formally combine the two anomalous loops into
the $\delta-$functional longitudinal susceptibility in which case we obtain the diagram 
in Fig.~\ref{Gorkeq}c. However, this is just a formal way to
reexpress the standard mean-field results.
Still, any inclusion of the 
renormalization of the internal fermionic line in fact means that one inserts
more pairs of anomalous loops into the self-energy, which one cannot do 
since this will render $\Sigma$ reduceable.

The Kampf-Schrieffer approach has the advantage over the conventional 
SDW decoupling in that it can straightforwardly be extended to the
region where long-range order is not present,
but where strong antiferromagnetic fluctuations exist, i.e. where
the correlation length is large, and the
spin susceptibility is strongly peaked, but not divergent,  
at zero frequency and at the antiferromagnetic momentum.
These are the basic assumptions for a nearly antiferromagnetic 
Fermi-liquid which we are going to study in the following section. 

\section{The nearly antiferromagnetic Fermi-Liquid}

\subsection{Transverse and longitudinal modes}

At first glance, the calculations performed in the ordered phase 
using the spin-fermion
model can be directly extended to the case of a 
``nearly'' $\delta$-functional form of the spin 
susceptibility in the disordered phase. However, this extension requires 
some care since in the
disordered phase, one no longer can distinguish between 
transverse and longitudinal fluctuations. Both modes behave in the same way, 
and both diverge as one approaches the antiferromagnetic
transition.
 
\begin{figure} [t]
\begin{center}
\leavevmode
\epsffile{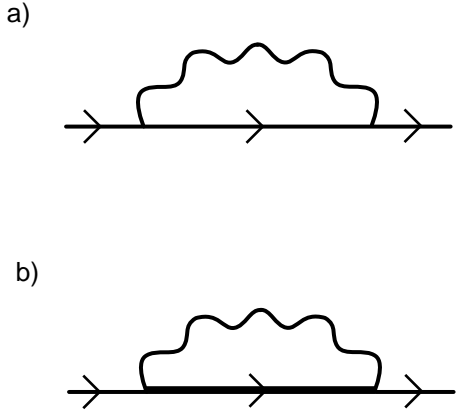}
\end{center}
\caption {{\em (a)} The lowest-order self-energy diagram for the
 spin-fermion model without long-range order. The wavy line describes 
 the exchange of spin fluctuations. All three fluctuation modes contribute equally to the 
 self-energy.
 {\em (b)} The self-energy diagram in the FLEX approximation. 
 The double line describes the full fermionic 
 propagator.}
\label{figse}
\end{figure}

 Now, if we apply the second-order diagram to evaluate
the Green's function (see Fig.\ref{figse}a), we do obtain the SDW form 
of the electronic spectrum 
as Kampf and Schrieffer have demonstrated, 
provided that the coupling is larger than the upper cutoff, $\omega_0$, 
in the spin susceptibility.
 However, because all
three modes (two transverse modes and one longitudinal mode) 
equally contribute to the
self-energy, the second-order self-energy at large coupling $g \gg \omega_0$ 
is given by
\begin{eqnarray}
\Sigma_{\alpha} (k,\omega) &=& g^2 ~
\sum_{a=1}^3 \sigma^a_{\alpha \beta} \sigma^a_{\beta \alpha} \int \chi_{aa}
(q,\Omega)G_0 (k+q, \omega + \Omega) \nonumber \\
&\approx& \frac{3}{4} ~g^2 ~G_0 (k+Q, \omega)
\label{dirkb}
\end{eqnarray}
 (we normalized $\int \chi_{aa} (q, \Omega)$ to $1/4$). 
This self-energy yields a ``near'' 
SDW solution with the relative corrections of
the order $(\omega_0/g)^2$, 
but with the gap larger by a factor of $\sqrt{3}$
than the one in the ordered phase (i.e., $\Delta^2 \approx 3 (g/2)^2$).  
At the same time, the vertex correction term at $g \gg \omega_0$ 
does not acquire an extra overall factor:
\begin{eqnarray}
\delta g /g
&=& g^2 ~
\sum_{a=1}^3 \sigma^a_{\alpha \gamma} 
\sigma^b_{\gamma \delta} \sigma^a_{\delta \beta}
 \int G_0(k+p, \omega+\bar{\omega} ) \nonumber \\
& & \hspace{-0.75cm} \times \; G_0(k+q+p, \omega +\Omega + \bar{\omega})
~\chi_{aa} (p,\bar{\omega}) \, dp \, d\bar{\omega}\nonumber \\
& & \hspace{-1.0cm} \approx -(g/2)^2~
G_0(k+Q,\omega) ~G_0(k+q+Q, \omega+\Omega)
\end{eqnarray}
As a result, the
relative vertex correction on the mass shell ($\omega \approx - \Delta, G_0 
\approx 1/ \Delta$) 
 will scale as $\delta g \approx - g~(g/2\Delta)^2$ and will only
account for a reduction of $g$ by a factor $1/3$. 
This in turn implies that the full vertex will not vanish for $q=Q$, contrary
to what one should expect if the mean-field SDW solution 
survives the loss of the sublattice magnetization. 
The reason for this discrepancy is that we have not yet included the
counterterm which in the ordered phase cancelled out the correction due to
transverse fluctuations  thus  eliminating two out of the three 
components of the spin susceptibility in the self-energy diagram.
Indeed, if we leave just one component of $\chi$ in Eq.(\ref{dirkb}), 
we find for the gap $\Delta\approx (g/2)$ in which case
$\delta g \approx - g$ and one recovers the correct form of the vertex.

\begin{figure} [t]
\begin{center}
\leavevmode
\epsffile{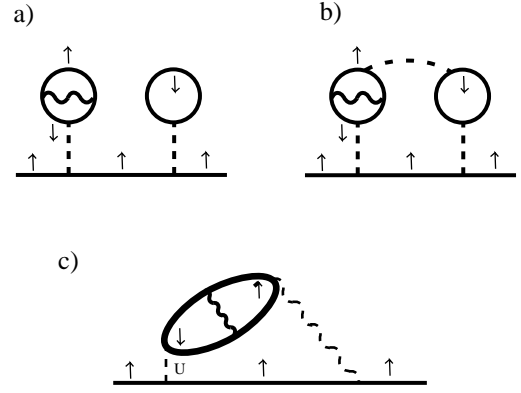}
\end{center}
\caption {The counter 
diagram which cancels the contribution from the lowest order self-energy 
term due to transverse spin fluctuations in the ordered state. The transformation 
from {\em a} to {\em c} is explained in the text. 
The solid and dashed lines represent the bare fermionic 
propagator and the interaction $U$, respectively. 
The solid and dashed wavy line describe 
transverse and longitudinal spin fluctuations, respectively. 
Observe that this diagram still contains the direct fermion-fermion 
interaction $U$.}
\label{counter}
\end{figure}

In order to identify the diagram which eliminates two components of the
spin susceptibility in the paramagnetic phase,
 we go back to the ordered state, reformulate the 
mean-field SDW theory as a result of the
exchange of spin fluctuations  and 
find the diagram which cancels the contribution from the 
transverse susceptibility.
 Obviously, this counterdiagram should include a
correction to $<S_z>$. The corresponding diagram for the Hubbard model
is shown in Fig.~\ref{counter}a.

 Note that the
internal lines in the anomalous loop are 
the full quasiparticle Green's functions given by Eq.(\ref{sdw-like}). 
Performing the same
manipulations that lead to the exchange of longitudinal fluctuations
(i.e., connecting the two anomalous bubbles by $U$ (Fig.\ref{counter}b)
and then substituting the single bubble by the full RPA series), 
we arrive at the diagram shown in Fig.~\ref{counter}c. 
This diagram includes spin-fermion vertices for the
exchange of both, transverse and longitudinal fluctuations, 
but also the bare Hubbard $U$.
The presence of the bare $U$ is crucial - it
implies that this diagram cannot be obtained in the 
pure spin-fermion model since it neglects a
 direct fermion-fermion interaction.
One can easily check that for a $\delta$-functional  
form of the longitudinal susceptibility
this diagram and the second-order diagram with transverse spin exchange
cancel each other in the large $U$ limit leaving the diagram 
with longitudinal spin exchange 
as the only relevant self-energy diagram. 

Suppose now that  the long-range order is lost but 
spin fluctuations are strong.
Then our reasoning is the following. 
Assume first that 
only one component of the susceptibility should be counted in the self-energy.
Then, as we just discussed, we obtain an SDW solution for $g=U \gg \omega_0$.
We then substitute the SDW form of $G(k,\omega)$ into the remaining part of 
the self-energy (the same as in Fig.\ref{figse}a, but with only two components of the 
susceptibility left) and into the 
fermionic bubble of the diagram in Fig.~\ref{counter}c.
Computing the total
contribution from these two diagrams we found that this contribution
is small compared to the one from
Fig.~\ref{Gorkeq}c by a factor $(\omega_0/g)^2$.
This in turn justifies the use of
only one component of the susceptibility in the second-order
self-energy.\\
Note that at small $g$ the situation is different: 
the counterdiagram in Fig.~\ref{counter}b
is of higher order in $g$ and can be neglected. In this
limit, we recover a conventional result of the lowest-order 
paramagnon theory that all three modes of spin fluctuations 
contribute to the self-energy.

\begin{figure} [t]
\begin{center}
\leavevmode
\epsffile{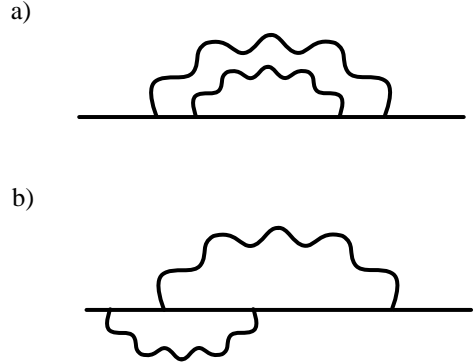}
\end{center}
\caption {Third order diagrams for {\em (a)} the self-energy and {\em (b)} 
the vertex correction. At large $g$ these two diagram nearly cancel each other. }
\label{third}
\end{figure}

We now  briefly discuss higher order corrections. In the paramagnetic phase,
they indeed exist and, moreover, 
are not small at large $g$.
However, extending the same line of reasoning as above 
to higher-order diagrams, 
one can easily verify that the 
higher-order self-energy and vertex corrections nearly cancel each other in the 
large $g$ limit. For example, the third-order self-energy and vertex correction 
terms are shown in
Fig.~\ref{third}.
 Assuming that just one fluctuation mode contributes to
the renormalization of the internal fermionic line, we find after simple
algebra that the two corrections nearly (to order $(\omega_0/g)^2$)
cancel each other. One, therefore can restrict with only the second-order
diagram and hence fully recovers the mean-field SDW solution.  
 
Before we conclude this section we want to comment on what happens if we restrict 
our calculations to the pure spin-fermion model. In this case each inclusion of the 
self-energy correction introduces a factor of 3 due to the absence of a 
counterterm. At the same time, the inclusion of the 
vertex correction only yields a relative factor of 
one. This implies that the self-energy corrections are more relevant
than the corrections to the vertex. 
This point can be made rigorous by extending the $SU(2)$ spins to $SU(N)$
and taking the limit of large $N$~\cite{CSS}.
For general $N$, one substitutes the Pauli matrices $\vec{\sigma}$ 
 in (\ref{intham})
by the $(N^2 -1)$ traceless generators of $SU(N)$. In the large $N$ limit, 
the vertex corrections are small by a factor of
$1/(N^2-1)$ and their relative contribution disappears at
 $ N \rightarrow \infty $.
At the same time, higher-order self-energy 
corrections do not contain extra powers of $1/N$
compared to the second-order diagram, and therefore should all be included. The
full self-energy correction is then given by a series of 
rainbow diagrams, which 
obviously can be reexpressed as a second-order diagram  with 
the full Green's function for the intermediate fermion (see Fig.\ref{figse}b). This approximation 
is known as the
fluctuation exchange (FLEX) approximation 
(the fully self-consistent FLEX approximation 
also uses the full fermionic Green's functions in the RPA series 
for the spin susceptibility).
The solution of the FLEX equations 
always yields a large, Luttinger-type Fermi 
surface with progressively 
decreasing quasiparticle residue as the spin-fermion coupling increases.  
We argue  that at least in the large coupling limit, 
this procedure is incomplete because for each inclusion of the rainbow diagram
one should also include a counterterm which effectively eliminates $N^2 -2$
components of the susceptibility out of the self-energy correction. Performing calculations along 
this line, one indeed recovers the SDW results.

\subsection{The model}
\label{toy}

We now discuss the model we are going to study. In a ``first principles''
calculation, we would have to
consider the Hubbard model. At large $U$, this model contains two peaks in the 
density of states (the upper and lower Hubbard band) for any doping concentration.
Far away from half filling, however, the total spectral weight of the upper band 
is small, and the excitations in this band are likely to be incoherent. As the 
system approaches half-filling, the spectral weight is shifted from the lower to 
the upper band and one gradually
recovers the SDW form of the spectrum even before the system becomes
magnetically ordered. How this evolution occurs is one of the key issues in 
understanding the normal state of the cuprates. We have not 
yet solved this problem in the Hubbard model, instead, we considered the evolution 
of the spectral function in a toy spin-fermion model. We assume that the density
of holes is fixed at some small but finite level, and vary
 the coupling constant $g$. In doing this, we in fact model the system's behavior
as it approaches half-filling simply because by all accounts, the spin-fermion
coupling should increase as the system becomes more ``magnetic''. 

Further, for any $g$, we compute the self-energy corrections restricting with just one
component of the spin susceptibility.  
The argument here is that, at strong
coupling, a counterterm which we discussed in the previous section, 
cancels out the corrections due
to the other two components of the susceptibility. At  small couplings, we indeed
will miss the overall factor of $3$ in the
self-energy, but this does not seem relevant as
we do not expect any qualitative changes in the fermionic 
spectrum. On the other hand, the study of
the toy model should give us the answers to two key questions: (i) how does the 
Fermi surface evolve from a large one, centered at $(\pi,\pi)$, at small
couplings to a 
small one, centered at $(\pi/2,\pi/2)$ at large $g$, 
and (ii) how does the 
spin-fermion vertex evolve from a momentum-independent one at small $g$
to one which vanishes at $Q$ (upto $(\omega_0/g)^2$) terms) at large
couplings.  

Finally, we assume that 
the  susceptibility has the form of 
Eq.(\ref{chi}) with some given $\omega_{sf}, ~\xi$ and $c_{sw}$ which do not
depend on the strength of the spin-fermion coupling.
This is indeed an approximation.
 When $g$ is large and precursors of the
SDW state are already formed,
 the particle-hole bubbles which constitute the RPA series
for the susceptibility 
should indeed be computed with the full Green's functions and the full vertices. However,
as we will see below, at large $g$, we in fact probe the scales larger
than $\omega_0$, in which case 
the structure of $\chi ({\bf q}, \omega)$ at $\omega < \omega_0$ is irrelevant
- we only use the fact that the susceptibility
obeys the sum rule. At weak coupling, $g < \omega_0$, the form of the
susceptibility is relevant, but in this limit the real part of the susceptibility 
is just an input function, independent on $g$, while the imaginary
part of $\chi$  can be computed  using the bare fermionic 
Green's functions and vertices which, in fact, we will do later.

The form of $\chi_{ii}({\bf q}, \omega)$ as in Eq.(\ref{chi}) 
is indeed an expansion near $\omega =0$ and $q=Q$. Meanwhile, 
$\chi (q,\omega)$ should indeed satisfy the sum rule $\int d\omega \, d^2{\bf q} 
\, \chi_{ii}(q,\omega) = 1/4$. To impose the sum rule, one should either add extra
terms into Eq.(\ref{chi}) with higher powers of $(q-Q)^2$, or impose a cutoff
in the momentum integration.  For computational purposes, it is easier to
impose a cutoff $\omega_0$. We have checked that
the results for the Fermi-surface evolution  practically do not depend on
whether we impose a cutoff only in the momentum integration or also in the
integration over frequency. In the latter case, however, the computations are
much easier to perform and a number of results can be obtained analytically.
Below we present the results for the model with a cutoff $\omega_0$
in both momentum and frequency. 

After presenting  the model, we now proceed to our calculations. We will
compute the full quasiparticle Green's function and the full vertex restricting with
the second-order diagrams only. Here we apply the same reasoning as before, namely that
in both the weak and the strong coupling limit, 
the relative corrections to the second-order expressions are small and 
scale as $(g/\omega_0)^2$ and $(\omega_0/g)^2$, respectively. This implies
that the second-order theory will yield a correct limiting behavior at weak and
strong couplings. At intermediate couplings, $g \sim \omega_0$, 
higher-order terms are indeed not small. However, even without them, 
we found in our numerical studies that for intermediate
$g$, the solutions for the Green's function and the vertex strongly depend
on the values of $\omega_{sf}, \xi$, etc., i.e.~the behavior is highly
nonuniversal. It is therefore unlikely
that the inclusion of higher-order diagrams will give rise to new
behavior which is not already present in the second-order theory.

We deem it essential to point out that, though we restrict the 
corrections to the 
second-order diagrams only, the computations of 
$G_{full} (k,\omega) = 
G_0 (k,\omega)/(1 - \Sigma (k,\omega) G_0 (k,\omega))$ 
($G_0$ is the fermionic Green's function at $g=0$)  and the full vertex 
contain a {\it nonperturbative}
 self-consistency procedure which is  necessary  to describe
a transformation from a large to a small Fermi surface. 
Namely,   we 
consider the chemical potential $\mu$ as 
an input parameter in the calculation of
the self-energy, and then determine it
 from the condition that the total density of
particles is equal to a given number, $\int d\omega d^2 k \, G_{full} (k, \omega)
= (1-x)/2$. The nonperturbative nature of this procedure appears since
the self-energy is  not small at large $g$, 
and one finds a finite
region in $(k,\omega)$ space where $\Sigma (k,\omega) G_0 (k, \omega) >1$
 i.e.~the perturbative expansion 
of $G_{full} (k,\omega)$  in powers of $\Sigma$ does not converge.  
This region, on the other hand, contributes to the density of particles, and
this makes the evaluation of $\mu$ a nonperturbative procedure.
 
We are now ready to present our results. We begin with
the self-energy corrections which give rise to the Fermi surface evolution.
Some of the results presented in the next subsection have already been reported 
earlier \cite{CMS}. 

\subsection{Fermi-surface evolution}

\begin{figure} [t]
\begin{center}
\leavevmode
\epsffile{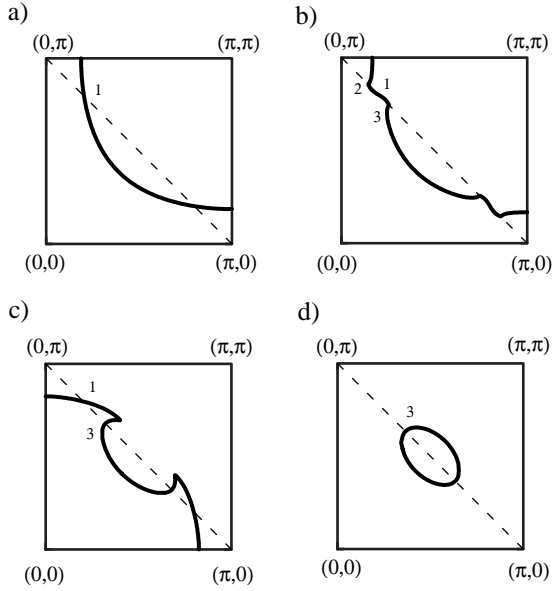}
\end{center}
\caption {The evolution of the hot spots with the coupling strength.
 For $g<g^{(1)}_{cr}$ (a) 
there exist only one hot spot; for $g=g^{(1)}_{cr}$ it splits into 
three (b). 
As  $g$ continues to increase, one progressively
looses two of the hot spots (c), and  for 
$g \gg g^{(2)}_{cr}$ (d) only a single ``new'' hot spot is present}
\label{exfs}
\end{figure}

\begin{figure} [t]
\begin{center}
\leavevmode
\epsffile{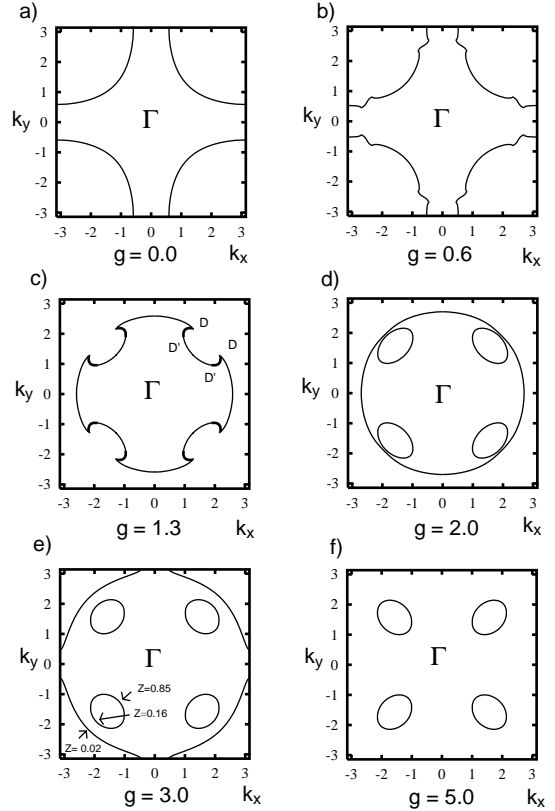}
\end{center}
\caption {Fermi surface evolution with increasing $g$.
 We used (in units of $t$), 
$\Delta_0=0.1$, $\omega_0=0.3$, $t^{\prime} =-0.45$, $x =0.1$. 
For these parameters we obtain 
$g^{(1)}_{cr} = 0.42$, and $g^{(2)}_{cr} \approx 1.64$.
The figures are for $g =0$, $2 \omega_0 > g > g^{(1)}_{cr}$ , 
 $g^{(2)}_{cr}> g >2 \omega_0$,
 $g \geq g^{(2)}_{cr}$, $g > g^{(2)}_{cr}$, 
and $g\gg g^{(2)}_{cr}$, respectively. In {\em (c)}, the
nested pieces of the Fermi surface
are shown in bold.  In Fig. {\em (e)}, we also presented the values
of the quasiparticle residue along the Fermi surface. Notice that the definition
of $g$ used in this paper differs by a factor of $2$ from the one in Ref.\protect\cite{CMS}.}
\label{fsevol1}
\end{figure}

As before, we assume that without a spin-fermion
interaction, the system behaves as a Fermi-liquid with a large Fermi surface
which crosses the magnetic Brillouin zone boundary (see Fig.\ref{exfs}a and \ref{fsevol1}a). 
Near the Fermi surface, the bare fermionic Green's 
function is $G_0 (k, \omega)  = Z/(\omega - (\epsilon_k - \mu))$,
where $\epsilon_k = -2 t
(\cos k_x + \cos k_y) - 4 t^{\prime} \cos k_x \cos k_y$.
The chemical potential at $g =0$ is clearly of the order of $t^{\prime}$:
$\mu = 4 t^{\prime} \cos^2 k^h_{x}$ where $k^h_x$ is the $x-$component of the
momentum for the point where the Fermi surface crosses the magnetic Brillouin
zone boundary. These points are generally known as "hot spots"~\cite{HlubRice}.  
In the Matsubara formalism, which is more convenient for computations, 
the self-energy at $T=0$ is given by
\begin{eqnarray}
\Sigma (k, \omega_m) &=& { g^2 \over (2 \pi)^3}\int d^2q \, d\Omega_m 
\chi (q,\Omega_m) \nonumber \\
& & \times \, G_0(k+q,\omega_m+\Omega_m)
\label{dirkc} 
\end{eqnarray}
As we discussed above, we will use the phenomenological form of the susceptibility
in Eq.(\ref{chi}) with a cutoff $\omega_0$ in both momentum and frequency.
The location of the Fermi surface is obtained from 
\begin{equation}
G^{-1}_{full}(k, \omega_m=0)= - \bar{\epsilon}_{k} - \Sigma (k, \omega_m=0)~=~0 
\label{fs1}
\end{equation}

It is instructive to consider first the case when the spin damping is absent, 
because then we can obtain a full analytical solution for $\Sigma$.
Since the susceptibility is peaked at $Q$, we can expand the energy of the
internal fermion  as
$\epsilon_{k+q} = \epsilon_{k+Q} + {\vec v}_{k+Q} ({\vec q}-{\vec Q})$.
Substituting this expansion into Eq.(\ref{dirkc}) and integrating over momentum
and frequency, we obtain  a rather complex function of the ratio 
$a_k = v_{k+Q}/ c_{sw} $ which we present in the Appendix.
The formula for $\Sigma (k, \omega_m)$ simplifies 
considerably if $a_k=1$. 
Then we have 
\begin{equation}
\Sigma (k, \omega_m) =  \frac{g^2}{4}~ {1 \over  i\omega_m -
 \bar{\epsilon}_{k+Q}} 
\label{aa}
\end{equation}
if $\bar{\epsilon}^2_{k+Q}+\omega_m^2 > \omega_0^2$ and
\begin{equation}
\Sigma (k, \omega_m) =  \frac{g^2}{4}~ {1 \over  i\omega_m -
 \bar{\epsilon}_{k+Q}}~\frac{\sqrt{\omega^2_m + 
\bar{\epsilon}^2_{k+Q}+ \Delta_0^2} - \Delta_0}{\sqrt{\omega^2_0  
+ \Delta_0^2} - \Delta_0}                       
\label{aaa}
\end{equation}
if $\bar{\epsilon}^2_{k+Q}+\omega_m^2 < \omega_0^2$.
Here we introduced $\Delta_0 = c_{sw}/\xi.$  

We have checked numerically that 
the qualitative picture for the evolution of the 
Fermi surface does not depend substantially on the value of $a_k$. 
We will therefore discuss the Fermi surface evolution by 
restricting with the form of the self-energy given in Eqs.(\ref{aa}) and (\ref{aaa}).

The self-energy in Eq.(\ref{aa}) is precisely what we need to recover the SDW
solution which, we recall, yields a small Fermi surface centered at
$(\pi/2,\pi/2)$. However, to obtain this small Fermi surface, Eq.(\ref{aa})
should be satisfied (at zero frequency) for all points in $k-$space, and, in
particular, for the points along the magnetic Brillouin zone boundary. 
At $g=0$, the  hot spots are at $\bar{\epsilon}_k = 
\bar{\epsilon}_{k+Q}=0$.  Clearly, at small $g$, 
the location of the Fermi surface near the
hot spots is obtained with the self-energy from Eq.(\ref{aaa}). A simple
experimentation shows that the location of the Fermi surface crossing does
not change with small enough $g$, i.e., it  still occurs at $\bar{\epsilon}_k = 
\bar{\epsilon}_{k+Q}=0$. However, as $g$ increases, there appear two
new hot spots at $\bar{\epsilon}_{k} \not = 0$. To see this, consider a point 
at the magnetic zone boundary right
near a hot spot. Expanding Eq.(\ref{aaa}) in $\bar{\epsilon}_{k+Q}$ and
substituting the result into Eq.(\ref{fs1}),  we obtain
\begin{equation}
G^{-1}_{full}(k, 0)= - \bar{\epsilon}_{k} ~\left(1 - \frac{g^2}{8 \Delta_0  \Big(
\sqrt{\omega^2_0 + \Delta_0^2} - \Delta_0 \Big) } \right)                         
\label{aaaa}
\end{equation}
For sufficiently small $g$, the 
only solution for the Fermi surface is $\bar{\epsilon}_{k} =0$.
However, when $g$ reaches the value 
$g_{cr}^{(1)}=2 \omega_0 [ 2\Delta_0/
(\sqrt{\omega_0^2+\Delta_0^2}+ 
\Delta_0)]^{1/2}$, the velocity along the magnetic zone boundary vanishes. 
For larger $g$, one still has a Fermi-surface crossing at 
 $\bar{\epsilon}_k=0$, however,  two new hot 
spots appear (see Fig.\ref{fsevol1}b) with 
\begin{equation}
\bar{\epsilon}_k=\pm { g^2 \over 4 \Big[ \sqrt{\omega_0^2+\Delta_0^2} - \Delta_0 \Big] } \Bigg( 1- 
{ ( g_{cr}^{(1)})^2 \over g^2 } \Bigg)^{1/2}
\label{nhs}
\end{equation}

As $g$ increases, $|\bar{\epsilon}_{k+Q}|$ for the new hot spots 
also increases, and
at $g = 2\omega_0$, these new hot spots satisfy 
$|\epsilon_{k+Q}| = \omega_0$, i.e., they become 
the solutions of $G^{-1}_{full} (k ,0) =0$ with the SDW-like form of the
self-energy in Eq.(\ref{aa}) rather
than the form in Eq.(\ref{aaa}). These solutions have a very simple form:
${\bar \epsilon}_k = \pm g/2$. For even larger $g$, 
the SDW form of the solution
for the Fermi surface extends beyond the hot spots and is 
 located
between points $D$ and $D^{\prime}$ in Fig.\ref{fsevol1}c.
As $g$ continues to increase, the $D$ and $D^{\prime}$ points from
neighboring hot spots approach each other. Finally, at 
$$g =g_{cr}^{(2)}= 
 2 \omega_0 \Bigg[ 1 +  { 8t(|\mu| -\omega_0) \over \omega_0 \Big[ 2t + \sqrt{4t^2 -
 4|t^{\prime}|(|\mu|-\omega_0)} \Big]  } \Bigg]^{1/2}$$
these points merge and the system undergoes a topological, 
Lifshitz-type phase transition (Fig.\ref{fsevol1}d) in which the 
singly-connected hole Fermi surface splits into hole pockets centered at 
$(\pm \pi/2, \pm \pi/2)$ and a large hole Fermi surface centered around 
$(\pi,\pi)$. As $g$ increases further, the large Fermi surface
shrinks (Fig.\ref{fsevol1}e) and eventually disappears at which point the
Fermi surface just consists of four hole pockets (Fig.\ref{fsevol1}f).

\begin{figure} [t]
\begin{center}
\leavevmode
\epsffile{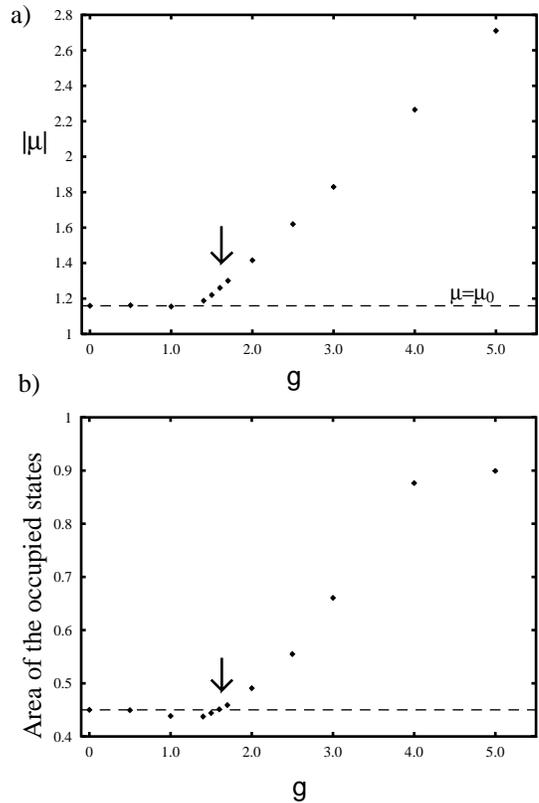}
\end{center}
\caption {{\em (a)} The chemical potential as a function of the coupling,
$g$. The parameters are the same as in Fig.~\protect\ref{fsevol1}.
For free fermions, $\mu =\mu_0 \approx -1.16$.
The arrow indicates the value of $g^{(2)}_{cr}$
when hole pockets are formed. {\em (b)} The area of the occupied states vs
$g$. The dashed line is the area for free fermions.}
\label{fsevol2}
\end{figure}

We found numerically
 that the  topological phase transition at $g_{cr}^{(2)}$ 
is accompanied by drastic changes in the 
functional behavior of the system. To demonstrate this point 
we plot in Fig.~\ref{fsevol2}a,b
 the chemical potential $\mu$ and 
the area of the electron states in the Brillouin zone as a function of 
$g$.
One clearly observes that both quantities 
are practically constant upto $g_{cr}^{(2)}$,
 but increase considerably for $g > g_{cr}^{(2)}$ 
(we attribute the small variations below $g_{cr}^{(2)}$ in Fig.~\ref{fsevol2}b
to numerical errors). 
In other words, Luttinger's theorem is satisfied below $g_{cr}^{(2)}$, but
is violated above this critical coupling.
We attribute the violation of Luttinger's theorem to the nonconvergence of
the perturbative expansion in $g$ above $g_{cr}^{(2)}$. We remind in this
regard that, in essence,  Luttinger's proof is perturbative: he showed
 that $I = \int G_{full} \partial \Sigma/\partial \omega=0$  order
by order in  perturbation theory. Implicit in this proof is the assumption
that the perturbative series is convergent. We computed $I$ numerically and
found that it is equal to zero (within the accuracy of our calculations)
below $g_{cr}^{(2)}$ but rapidly increases as soon as $g$ exceeds the
critical value.  We reserve a detailed discussion of
 this issue for a separate publication.

Further, we mentioned in the 
introduction that  it is still a subject of controversy whether 
the sides of the hole pockets which 
are facing $(\pi,\pi)$ have been observed in experiments or not.
The relevant physical quantity  here is the quasiparticle 
residue at the Fermi surface. 
For small $g$, when one has 
a large Fermi surface, $Z_k$ is almost $k-$independent and close to one. 
For very 
large $g$, when one effectively recovers the SDW form 
of the electronic excitations, 
$Z_k$ is again $k-$ independent and is equal to  $1/2$. 
For intermediate $g$ , however, $Z_k$ 
is strongly $k-$dependent and  anisotropic. The full expression for $Z_k$ is
presented in the Appendix. The results for the case when hole pockets 
and a large Fermi surface  coexist are presented (for $a_k =1$) in 
Fig.\ref{fsevol1}e. We see that the quasiparticle residue along the large Fermi
surface is very small, which makes the experimental observation practically
impossible. Furthermore, we see that the quasiparticle residue of the side of
the hole pocket which is facing $(\pi,\pi)$ is roughly five times smaller than 
the one in the momentum region facing the $\Gamma$ point. This may very well explain
the experimental difficulties in observing the 
part of the hole pocket facing $(\pi,\pi)$.           
 
Finally, we want to 
 discuss how the above results change when we take a damping of spin
fluctuations $\gamma = c^2_{sw}/(2 \omega_{sf} \xi^2)$ into consideration.
 We found that the general scenario of the Fermi surface evolution
does not change, but that its onset occurs at
substantially larger values of the coupling constant 
than in the absence of damping. Specifically, we found 
 (still, assuming for simplicity that 
$a_k =1$) that the value of $g^{(1)}_{cr}$ where a
hot spot splits into three is  $g^{(1)}_{cr} (\gamma) =
g^{(1)}_{cr} (\gamma =0) \Psi (\gamma/\Delta_0)$ where
$\Psi (x) = 1 - x/(3\pi) +O(x^2)$ for $x \ll 1$, and $\Psi (x) = (16 \log
x/(\pi x))^{1/2}$ for $x \gg 1$.  In the nearly antiferromagnetic Fermi liquid,
$\gamma/\Delta_0 = c_{sw}/(2 \omega_{sf} \xi)$ is a large parameter.
 In this case, $g^{(1)}_{cr} \propto
[\gamma /\log (\gamma/\Delta_0)]^{1/2}$
 which is substantially larger than $g^{(1)}_{cr}
\propto (\Delta_0)^{1/2}$ which we obtained in the absence of damping.

\subsection{Vertex corrections}
\label{vert}
                              
The bare interaction vertex in the spin-fermion model is a
momentum-independent constant $g$. 
At small couplings, the full vertex is indeed close to the
bare one. At strong couplings, on the other hand, we have precursors 
of the SDW state, and, as we discussed, 
the full vertex should be much smaller than the bare one.
 We now study how the full vertex evolves with the coupling strength. 
 Contrary to
naive expectations, we find that the evolution of the vertex is not smooth.  
For definiteness, we will study below
the vertex which describes the interaction between fermions at
 the hot spots and 
transverse spin fluctuations  with  
antiferromagnetic momentum $Q$ and zero frequency. 
This vertex is relevant to the pairing problem
since the incoming and outgoing fermions can simultaneously be placed
on the Fermi surface. 
  
\begin{figure} [t]
\begin{center}
\leavevmode
\epsffile{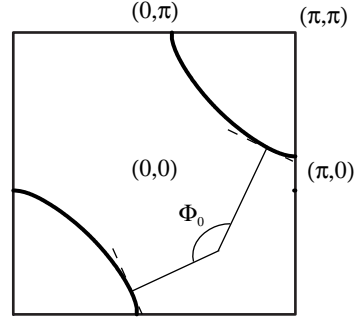}
\end{center}
\caption {The graphical representation of the angle $\phi_0$ 
between the normals to the Fermi surface 
at hot spots (dashed lines). 
For clarification we omitted the parts of the Fermi surface in 
the second and fourth quadrant.}
\label{phi0}
\end{figure}

Consider first the limit of weak coupling. 
The diagram for the vertex correction is presented
in Fig.~\ref{figsovert}. Since the spin 
susceptibility is peaked at $Q$, the internal fermions are also located near the hot 
spots, and all three internal lines in
 the diagram carry low-energy excitations. 
A simple dimensional
analysis shows that the vertex correction is logarithmically
singular in the limit when the external fermionic frequencies are zero and
the correlation length is large: $\Delta g \propto g * \log[max(\omega_{1,2}, 
c_{sw} \xi^{-1})]$.
In order to obtain an analytical 
expression for $\Delta g$, we expand the fermionic energies near the hot spots as
 $\epsilon_k = v_h(k-k_{h}) \cos (\phi)$ and
$\epsilon_{k+Q} = v_h(k-k_{h}) \cos (\phi + \phi_0)$ and
 perform the integration over frequency and
momentum in the spin susceptibility. Keeping both, the logarithmically divergent and 
the $\xi$-independent terms in $\delta g$,
 we obtain for zero external fermionic
frequency
\begin{eqnarray}
\frac{\Delta g}{g} &=& - \frac{g^2 Z^2\chi_Q \omega_{sf}}{\pi^3 v_h^2}
  Re \Bigg\{ \int_0^{\pi}~d\phi~
\frac{\log[\sin (\phi/2)]}{\cos \phi + \cos \phi_0}  \nonumber \\
& &\quad \times \log\frac{\sin(\phi/2)}{\delta^2} \Bigg\} + O(\delta^2) 
\label{beta}
\end{eqnarray}
where $\delta = c^2_{sw}/(2\gamma~v_h~\xi) \equiv
 \xi \omega_{sf}/v_h \ll 1$. 
 $Z$ is the fermionic 
quasiparticle residue in the absence of  the spin-fermion interaction and
 $\phi_0$ is the angle between the normals to the Fermi surface 
at $k$ and $k+Q$ (see Fig.~\ref{phi0}).
 For the $t-t^{\prime}$ model for the
quasiparticle dispersion  we find 
$\phi_0 = \pi/2 + 2
tan^{-1} [(1 - 2|t^{\prime}| \sqrt{\mu/4 t^{\prime}})/((1 + 2|t^{\prime}| \sqrt{\mu/4
t^{\prime}})]$. 
Observe that $\Delta g$ does not depend on the upper
cutoff in the frequency and momentum integration (i.e., on $\omega_0$).

We see that $\Delta g$ diverges logarithmically 
when $\xi$ diverges and $\delta \propto \xi^{-1}$
 tends to zero. Alternatively, one can compute $\Delta g$ 
at $\xi^{-1} =0$ but at finite $\omega_{1,2}$, and also obtain a logarithmical
divergence. This last result was also obtained by Altshuler {\it et al.}\cite{AIM}.

The integral in the right hand side 
of Eq.(\ref{beta}) can easily be evaluated numerically 
(and also analytically for particular values of $\phi_0$).  It turns out that this integral
is negative for all $\phi_0$, i.e., the relative vertex correction is {\it positive}. This
implies that for small couplings, vertex corrections in fact increase the coupling strength
and therefore enhance the $d-$wave pairing interaction. This is the opposite behavior 
of what we would expect at strong couplings. 
The strength of the vertex correction is another issue which
recently was the subject of some controversy~\cite{AIM,Chu,Stamp}.
 The key issue here is whether one 
should consider $\omega_{sf}$  as an independent input parameter, 
or assume that the damping is mainly due to the interaction with fermions.  In
the latter case, which we believe is closer to reality, 
$\omega^{-1}_{sf}$ by itself is proportional to $g^2 Z^2/v_h^2$
 in which case the 
coupling strength, fermionic velocity,
 and the bare quasiparticle residue are cancelled out in
 the r.h.s.~of Eq.(\ref{beta}).  
In explicit form we found for $\omega_{sf}$ 
\begin{equation}
\omega_{sf} = \frac{\pi}{4}~|\sin \phi_0|~ \frac{v_h^2}{g^2 Z^2 \chi_Q}.
\label{osf}
\end{equation}
Substituting $\omega_{sf}$ into
Eq.(\ref{beta}) we obtain, neglecting terms of $O(\delta^2)$
\begin{equation}
\frac{\Delta g}{g} = - \frac{|\sin \phi_0|}{4\pi^2}
Re \int_0^{\pi}~d\phi~
\frac{\log[\sin (\phi/2)]}{\cos \phi + \cos \phi_0} \log\frac{\sin
(\phi/2)}{\delta^2}
\label{beta2}
\end{equation}
This expression only depends on $\phi_0$.
Numerically, the r.h.s. of Eq.(\ref{beta2}) 
is small for all realistic values of $\delta$.
Thus, for $\delta =0.27$, $t^{\prime} =-0.45 t$, and $\mu =-1.16$ 
(which corresponds to 10\%
doping), we have $\Delta g/g \approx + 0.04$. 
A similar result has been obtained in Ref.\cite{AIM}.

Notice that though  the vertex correction is small numerically, 
Eq.(\ref{beta2}) shows that
the perturbative expansion for  $\Delta g$ breaks down since 
$\Delta g$ exhibits a step-like behavior when $g$ becomes finite. 
This behavior of $\Delta g$, however, is just an artifact of our approximation 
in which we neglected the $\omega^2$ term compared to $\omega/\omega_{sf}$
in the spin susceptibility.
 A more detailed analysis shows
that a continuous behavior of $\Delta g$ is restored for $g < \delta$.

Consider now what happens when $g$ increases and the Fermi surface 
starts to evolve. As we
discussed in the previous subsection, the evolution begins at 
$g = g^{(1)}_{cr}$
 with the flattening of the Fermi
surface at the hot spots. For $g > g^{(1)}_{cr}$ there appear 
 two satellites of the original hot spot (see Fig.~\ref{exfs}b).
 The central hot spot is a solution of $\epsilon_k = \mu$  below and above
$g^{(1)}_{cr}$, and the vertex corrections for this hot spot are virtually insensitive to the
onset of the Fermi surface evolution 
(we recall that we restrict ourselves with the second-order
correction which involves only bare fermionic propagators). 
On the other hand, for the
two new hot spots, $\epsilon_k - \mu$ is finite (see Eq.(\ref{nhs})), 
and thus the vertex 
corrections will be different. 
As the Fermi surface continues to evolve, the old hot spot and
one of the two new hot spots eventually disappear leaving a single new hot spot as the 
only remaining one on the Fermi surface. 
The vertex correction at this hot spot depends
on the value of the chemical potential and changes sign when $|\mu|$,
 which grows with $g$ for  $g > g^{(2)}_{cr}$,
 becomes comparable to the fermionic bandwidth.
 At even larger $g$, $|\mu|$ grows
approximately as $g/2$. In this limit, 
the relative vertex correction becomes, to
leading order in $\omega_0/g$, $\Delta g/g \approx - (g/\mu)^2 
\int d^2 q d\Omega \chi_{aa} (q,\Omega) \approx -1$, i.e., 
the total vertex $g_{tot} = g + \Delta g$ is
nearly zero. Performing calculations beyond the leading order in $\omega_0/g$
and for spin momentum $q \neq Q$,  we found that in this limit, 
\begin{equation}
g_{tot} (q) \propto \frac{\omega_0}{g}~~\left(  \lambda^2 
\left(\frac{\omega_0}{g}\right)^2 + (q-Q)^2 \right )^{1/2} ,  
\label{tot}
\end{equation}
where $\lambda=O(1)$ is a numerical factor.
We see that for very large couplings, the vertex is nearly
linear in $(q-Q)$ as it should be in the SDW phase. 

The strong coupling form of the vertex
is very similar to the one suggested by Schrieffer~\cite{Schr}, 
the only difference is that in his
expression, $(\omega_0/g)^2$ in Eq.(\ref{tot}) is replaced by $\xi^{-2}$. We argue
that one needs more than just a large correlation length to recover the precursors of the SDW
state in the electronic dispersion and the vertex. Namely, in our approach, 
the SDW form of the
vertex appears due to a separation of the original fermionic dispersion 
into two subbands
separated by an energy scale $g$  which is large compared to 
the cutoff $\omega_0$ in the spin
susceptibility. When the correlation length becomes 
larger, $\omega_0$ clearly goes
down (and hence, $g/\omega_0$ increases), but 
$\omega_0$ remains finite when $\xi$ becomes
infinite. When the system becomes 
ordered, there appears a new 
momentum scale, $q_0 \propto <S_z>$, such that when $|q-Q| <q_0$, 
$g_{tot} (q)$ is
strictly linear in $|q-Q|$ in accordance with the Adler principle,
 while for larger momentum,
one obtains a crossover to Eq.(\ref{tot}). 
For $q_0 \gg (\omega_0/g)$, the difference 
between the two limits becomes negligible and one fully recovers the strong-coupling 
SDW result for the vertex.

\section{Conclusions}
\label{concl}
We first summarize our results. The goal of 
this paper was to study the evolution of  the
electronic and magnetic properties of cuprates as one moves 
from optimal doping into the
underdoped regime. Specifically, we were mostly interested in a 
possible relation between
the reduction of $T_c$ with decreasing doping and the loss of the 
pieces of the Fermi surface
near $(0,\pi)$ and related points, as evidenced in photoemission data. 
To address this issue, we considered an effective spin-fermion 
model in which itinerant fermions interact with low-energy spin fluctuations 
whose dynamics are described by a
semi-phenomenological spin susceptibility which we assume to be peaked at the
antiferromagnetic momentum. 
In the absence of the spin-fermion interaction, the fermions are assumed to 
form a Fermi liquid with a large Fermi surface which encloses an 
area roughly equal to  half of the Brillouin zone. 
We argued that one can model the system's evolution towards half-filling by increasing the
strength of the spin-fermion interaction. The weak coupling limit 
models the situation
near optimal doping, while the strong coupling limit corresponds to 
strongly underdoped cuprates. We found that, as the coupling increases, the Fermi surface first
evolves in a continuous way, and its area remains unchanged. At some critical coupling,
however, the system undergoes a topological, Lifshitz-type phase transition in which
the singly connected hole Fermi surface centered at $(\pi,\pi)$ 
splits into a hole pocket centered at $(\pi/2,\pi/2)$, and a large hole Fermi surface
 centered at $(\pi,\pi)$. As $g$ increases further, the area of the 
large Fermi surface gradually decreases, and it eventually disappears.
 As a result, the Fermi
surface at large enough $g$ consists of four small pockets. 
Simultaneously with the changes
in the Fermi surface topology, the spin-fermion vertex also 
changes from being nearly 
constant at small $g$ to being nearly a linear function of $q-Q$ at strong couplings. 
This last form of the vertex is the one expected for an ordered
antiferromagnet. When the vertex is linear in $q-Q$, one 
does not obtain an attractive
interaction in the $d-$wave pairing channel since the enhancement of the spin susceptibility
near $Q$ is fully compensated by the reduction of the vertex. 

Our results, therefore, show that there is an interplay
 between the Fermi surface evolution and
the loss of $d-$wave superconductivity in strongly underdoped cuprates. 
Both phenomena are
related to the fact that the strong coupling SDW forms of the electronic dispersion and
the spin-fermion vertex do not change drastically when the system looses long-range magnetic
order, contrary to what one could expect from the mean-field theory.
Instead, the electronic structure gradually changes towards a conventional Fermi liquid 
as magnetic correlations become less and less relevant. We believe
that the Fermi liquid
picture of electronic states, with one single peak in the density of states, 
is restored
somewhere around optimal doping. 

As we already mentioned in the introduction, 
the transformation from a large to a small
Fermi surface in underdoped cuprates is consistent with  
the experimental data, though
only one group so far reported the observation of both sides of 
the hole pocket. We
emphasize, however, that in the intermediate coupling regime, 
which most probably corresponds
to the experimental situation for $T_c = 30K$ and $T_c =60K$ $Bi2212$ 
superconductors 
studied by photoemission~\cite{Shen,LaRosa,Ding}, 
the quasiparticle residue for the part of the pocket which faces
$(\pi,\pi)$ is rather small (few times smaller than on the other side of 
the pocket) which
makes it more difficult to extract the quasiparticle peak 
from the background. 

There are several issues which are not resolved in this paper and 
require further study.
First, the photoemission data indicate that 
the quasiparticle peak in the spectral function
is broad already at optimal doping and becomes even broader as the system moves
towards half-filling. We have shown in Sec.~\ref{fleff}
 that at exactly half-filling, the
broadening of the quasiparticle dispersion is due to the
interaction with {\it short wavelength}
spin fluctuations 
(long wavelength fluctuations do not contribute because of the vanishing
vertex).
 In our analysis of the spin-fermion model with a cutoff $\omega_0$ in  the 
susceptibility,
 we completely neglected these fluctuations in the paramagnetic phase
 (we remind that
in our discussion on the cancellation of the higher-order 
self-energy diagrams, we restricted
with only the leading terms in the expansion in the bosonic frequency and 
the momentum shift from $Q$. We do not believe that
the qualitative features of the Fermi-surface evolution will 
change if we include short-wavelength spin
fluctuations, but they can be important for a quantitative analysis. 
For example, the FLEX
calculations, which do not yield precursors of the SDW state, 
also show that $T_c$ is reduced
in underdoped cuprates simply because fermions become less coherent 
as one approaches
half-filing~\cite{berlin}.

Another issue which requires further study has emerged from
the photoemission data from the Stanford and Argonne groups~\cite{Shen,Ding} 
which found  that, 
in  underdoped cuprates, $A(\omega)$ for $k$ near $(0,\pi)$  
not only has a broad
peak at about $2J$, as it should be if the precursors of 
the SDW state are present, but
also drops rapidly at frequencies of about $30 meV$. 
There is no such drop for the data
taken along the Brillouin zone diagonal. 
When the temperature is lowered below $T_c$, the
spectral function acquires a narrow peak at exactly the same 
position where the drop  has been observed~\cite{comm}. 
It is tempting to associate this new feature with the precursors
of the $d-$wave pairing state. Emery and Kivelson proposed
a scenario for $d-$wave precursors 
in which proximity to the antiferromagnetic instability
is not particularly relevant~\cite{EK}. A similar scenario has been suggested by
Randeria {\it et al.} \cite{Rand}.
 From our perspective, an important point is the observation 
by Shen {\it et al.}~\cite{Sh}
that both, the high energy peak and the drop in $A(\omega)$ at low energies 
disappear at about the
same doping concentration, i.e., the two features are 
likely to be correlated. This
observation poses the question whether precursors of the SDW state can give 
rise to
precursors of the $d-$wave pairing state. At the moment, we do not know the answer to this
interesting question.

Finally, the reasons why Luttinger's theorem does not work above the critical
value of the coupling are still not completely clear to us. Recently, we
considered in detail why Luttinger's proof for $I = \int G_{full} \partial 
\Sigma /\partial \omega =0$ does not work in the
magnetically-ordered state~\cite{Luttord}. We found that though a formal
application of Luttinger's arguments yields $I=0$ to all orders in the SDW gap $\Delta$ in
perturbation theory, the frequency integrals contain a
hidden linear divergence and have to be regularized. When this regularization
is done, one obtains that $I$ is finite, as it should be to recover Eq.(\ref{area}), and 
for small $\Delta$ behaves as $I \propto \Delta^2$. We are currently studying whether the
same reasoning can be applied above the topological transition in the
paramagnetic phase.
  
We conclude with a final remark. 
In this paper, we studied the Fermi surface evolution 
as a function of the
coupling strength at zero temperature. There exist, however, a number of experimental data
which show temperature crossovers in various observables in the underdoped cuprates, most
noticeably in the NMR relaxation rate, the uniform susceptibility and the resistivity.
 Recently, Pines, 
Stoikovich and one of us (A. Ch.) argued that these crossovers are related to the thermal
evolution of the Fermi surface (or, more accurately, to the evolution of the quasiparticle peak
in the spectral function)~\cite{CPS}.
 We refer the interested reader to Ref.\cite{CPS} for a detailed
discussion of this issue.

\section{acknowledgements}
It is our pleasure to thank 
all colleagues with whom we discussed the issues considered in
this paper. The research was supported  by NSF DMR-9629839.
A. Ch. is a Sloan fellow.

\appendix
\section{The self-energy for arbitrary $\lowercase{a}_{\lowercase{k}}$}
The explicit computation of Eq.(\ref{dirkc}) yields for arbitrary 
$a_k=|v_{k+Q}/ c_{sw}|$
\begin{eqnarray}
\Sigma (k, \omega_m) &=& - g^2_{eff} { \sqrt{(\bar{\epsilon}_{k+Q} - i
\omega_m)^2} \over  \bar{\epsilon}_{k+Q} - i\omega_m} \nonumber \\
& &\times \; {1 \over \Big(\sqrt{\omega_0^2+
\Delta_0^2} -\Delta_0 \Big) \sqrt{1-a_k^2} } \times A
\label{sigma1} 
\end{eqnarray}
where
$$
\Delta_0 = c_{sw}/\xi,\; g_{eff}^2=g^2 \alpha 
{\sqrt{\omega_0^2+\Delta_0^2}-\Delta_0 \over \pi}  
$$
and 
\begin{equation}
A = \ln{ {\sqrt{(\bar{\epsilon}_{k+Q} - i\omega_m)^2+\omega_0^2(1-a_k^2)} + 
\sqrt{(1-a_k^2)(\omega_0^2+\Delta_0^2)} \over \sqrt{(\bar{\epsilon}_{k+Q} - i
\omega_m)^2} +\Delta_0 \sqrt{1-a_k^2} } }
\label{sigma2}
\end{equation}
if $\bar{\epsilon}_{k+Q}+\omega_m^2 f_a > a_k^2\omega_0^2$,
 and
\begin{eqnarray}
A &=& \ln \Bigg[ {\sqrt{(\bar{\epsilon}_{k+Q} - i\omega_m)^2+\bar{\epsilon}_{k+Q}+
\omega_m^2~f_a (1-a_k^2)} \over a \Big( \sqrt{(\bar{\epsilon}_{k+Q} - i
\omega_m)^2} +\Delta_0 \sqrt{1-a_k^2} \Big) } + \nonumber \\
& & + { \sqrt{(1-a_k^2)(\bar{\epsilon}_{k+Q}+\omega_m^2~f_a + a_k^2\omega_0^2)}
\over a_k \Big( \sqrt{(\bar{\epsilon}_{k+Q} - i
\omega_m)^2} +\Delta_0 \sqrt{1-a_k^2} \Big) } \Bigg]
\label{sigma3}
\end{eqnarray}
if $\bar{\epsilon}_{k+Q}+\omega_m^2 f_a > a_k^2\omega_0^2$ with
$$   
f_a={ (1+a_k)^2-3 \over a_k}   
$$
For $a_k=1$ we recover the expressions presented in Eqs.(\ref{aa}) and (\ref{aaa}).\\
We also can extract the quasiparticle residue at the Fermi surface.
Substituting (\ref{sigma1}) into 
$$ {1 \over Z_k} = 1- {\partial \Sigma \over \partial \omega}\Bigg|_{\omega=\bar{\epsilon}_k}$$
we 
we obtain
\begin{equation}
{1 \over Z_k} = 1+{ g_{eff}^2 \; B \over \Big(\sqrt{\omega_0^2+\Delta_0^2} -
\Delta_0 \Big) 
\Big( |\bar{\epsilon}_{k+Q}|+\Delta_0 \sqrt{1-a_k^2} \Big) } 
\end{equation}
where
\begin{equation}
B= { \sqrt{ \bar{\epsilon}_{k+Q}^2 +\Delta_0^2a_k^2}-a_k^2\Delta_0 + 
|\bar{\epsilon}_{k+Q}| \sqrt{1-a_k^2} \over  |\bar{\epsilon}_{k+Q}|+ 
\sqrt{1-a_k^2} \sqrt{  \bar{\epsilon}_{k+Q}^2 +\Delta_0^2a_k^2 } } 
\end{equation}
if $|\bar{\epsilon}_{k+Q}|<\omega_0a_k$ and
\begin{eqnarray}
B &=& \Bigg\{ \sqrt{ \omega_0^2 +\Delta_)^2} + { \omega_0^2\sqrt{ 1-a_k^2}-
\Delta_0|\bar{\epsilon}_{k+Q}| \over \sqrt{\bar{\epsilon}_{k+Q}^2 +\omega_0^2(1-a_k^2)}} 
\Bigg\} \nonumber \\
&\times& {1 \over   
\sqrt{1-a_k^2} \sqrt{ \omega_0^2 +\Delta_0^2} + \sqrt{\bar{\epsilon}_{k+Q}^2 +
\omega_0^2(1-a_k^2)} }
\end{eqnarray}
if $|\bar{\epsilon}_{k+Q}|>\omega_0a_k$.   We plotted this result (for $a_k=1$)
in Fig.~\ref{fsevol1}e.


\begin{references}
\bibitem{comm1}  There are two nonequivalent $Cu$ positions in the unit cell of
$Bi2212$ compounds which by itself can generate $(\pi,\pi)$ modulations.
\bibitem{comm5} Indeed, very near $q=Q$, $g_{eff}$ is linear in $|q-Q|$ for any 
coupling. In the weak coupling limit, however, $g_{eff}$ quickly saturates to its 
bare value.
\bibitem{comm2}We do not consider here 
possible exponential tails in the density of states (Lifshitz tails).
\bibitem{comm}By itself, this result, if interpreted 
in the framework of the conventional
BCS theory,  implies that the density of particles which 
contribute to the coherent peak at the superconducting gap 
is much smaller than the total
density of fermions.
\end{references}
\end{document}